\documentclass[]{aastex}
\usepackage{emulateapj5}
\usepackage[figuresright]{rotating}
\usepackage{natbib}

\newcommand{\sz}{$S_z$}
\newcommand{\exo}{{\it EXOSAT}}
\newcommand{\gin}{{\it Ginga}}
\newcommand{\xte}{{\it RXTE}}
\newcommand{\mdot}{$\dot{M}$}

\begin{document}

\title{{\it RXTE} observations of the neutron star low-mass X-ray
binary \\ GX 17+2: correlated X-ray spectral and timing behavior}

\author{Jeroen Homan\altaffilmark{1,8},
        Michiel van der Klis\altaffilmark{1}, 
	Peter G. Jonker\altaffilmark{1}, 
	Rudy Wijnands\altaffilmark{2,3}, 
	Erik Kuulkers\altaffilmark{4,5}, 
	Mariano M\'endez\altaffilmark{6}, 
	\& Walter H. G. Lewin\altaffilmark{7}} 

\altaffiltext{1}{Astronomical Institute 'Anton Pannekoek', University
of Amsterdam, and Center for High Energy Astrophysics, Kruislaan 403,
1098 SJ Amsterdam, The Netherlands; homan@astro.uva.nl,
michiel@astro.uva.nl, peterj@astro.uva.nl}

\altaffiltext{2}{Center for Space Research, MIT, 77 Massachusetts
Avenue, Cambridge, MA 02139-4307, USA; rudy@space.mit.edu}

\altaffiltext{3}{Chandra fellow} 

\altaffiltext{4}{Space Research Organization Netherlands,
Sorbonnelaan 2, 3584 CA Utrecht, The Netherlands; E.Kuulkers@sron.nl}

\altaffiltext{5}{Astronomical Institute, Utrecht University, P.O. Box
80000, 3508 TA Utrecht, The Netherlands}

\altaffiltext{6}{Facultad de Ciencias Astron\'omicas y
Geof\'{\i}sicas, Universidad Nacional de La Plata, Paseo del Bosquey
S/N, 1900 La Plata, Argentina; M.Mendez@sron.nl}

\altaffiltext{7}{Department of Physics and Center for Space Research,
MIT, 77 Massachusetts Avenue, Cambridge, MA 02139-4307, USA;
lewin@space.mit.edu}

\altaffiltext{8}{Current address: Osservatorio Astronomico di Brera, Via E. Bianchi 46, 23807 Merate LC, Italy; homan@merate.mi.astro.it}

\begin{abstract} We have analyzed $\sim$600 ks of {\it Rossi X-ray
Timing Explorer} data of the neutron star low-mass X-ray binary and Z
source GX 17+2. A study was performed of the properties of the noise
components and quasi-periodic oscillations (QPOs) as a function of
the broad-band spectral properties, with the main goal to study the
relation between the frequencies of the horizontal branch (HBO) and
upper kHz QPOs. It was found that when the upper kHz QPO frequency is
below 1030 Hz these frequencies correlate, whereas above 1030 Hz they
anti-correlate. GX 17+2 is the first source in which this is
observed. We also found that the frequency difference of the high
frequency QPOs was not  constant and that the quality factors (Q
values) of the HBO, its second harmonic, and the kHz QPOs are
similar, and vary almost hand in hand by a factor of more than three.
Observations of the normal branch oscillations during two type I
X-ray bursts showed that their absolute amplitude decreased as the
flux from the neutron star became stronger. We discuss these and
other findings in terms of models that have been proposed for these
phenomena. We also compare the behavior of GX 17+2 and other Z
sources with that of black hole sources and consider the possibility
that the mass accretion rate might not be the driving force behind
all spectral and variability changes.\end{abstract}


\section{Introduction} 

Based on their broad-band spectral and variability properties, six of
the persistently bright neutron star low-mass X-ray binaries (LMXBs)
were classified as Z sources \citep{hava1989}, after the Z-like
tracks they trace out in X-ray color-color (CDs) and
hardness-intensity diagrams (HIDs). These sources are GX 17+2, Cyg
X-2, GX 5-1, GX 340+0, Sco X-1, and GX 349+2.  The Z-like tracks
consist of three branches, which, from top to bottom, are referred to
as the horizontal branch (HB), the normal branch (NB), and the
flaring branch (FB). It is generally
believed that the parameter that determines the position along the Z
track is the mass accretion rate, increasing from the HB to the FB.
Note that the definition of the mass accretion rate in this respect
is rather vague -- it is often defined as the mass accretion rate
through the inner disk and onto the neutron star surface. In addition
to spectral changes along the Z-track, some of the Z sources show
long term changes in the shape and position of the Z-track in the CD
and HID. These secular changes, as they are referred to, are clearly
observed in Cyg X-2 \citep{kuvava1996,wivaku1997}, GX 5-1
\citep{kuvaoo1994}, and GX 340+0 \citep{kuva1996} (the Cyg-like
sources; the other Z sources are referred to as the Sco-like
sources), and more recently also in GX 17+2 \citep{wihova1997}. It
has been suggested that they are related to the relatively high
inclination at which these sources are seen
\citep{kuvaoo1994,kuva1995}, or to different magnetic field strengths
of the neutron stars \citep{pslami1995}. 

The power spectra of the Z sources show several types of
quasi-periodic oscillations (QPOs) and noise components
\citep[see][for a review]{va1995a}.  It was found that their presence
and properties are very well correlated with the position of the
source along the Z track \citep{hava1989}, even when the Z-tracks
show secular changes \citep[e.g.][]{kuvaoo1994}.  Three types of low
frequency ($<$100 Hz) QPOs are seen in the Z sources: the horizontal
branch (HBOs), normal branch (NBOs) and flaring branch QPOs (FBOs).
Their names derive from the branches on which they were originally
found. The HBO is found on the HB and NB with a frequency (15--60 Hz)
that gradually increases along the HB towards the NB. When the
sources move from the HB onto the NB the frequency increase flattens
off. In GX 17+2 and Cyg X-2 it was found that when the source passes
a certain point on the NB, the HBO frequency starts to decrease 
\citep{wivaps1996,wivaku1997}. The NBO and FBO are most likely the
same phenomenon. They are found on the NB and FB (near the NB/FB
vertex) but not on the HB. On the NB the QPO has a frequency of
$\sim$5--7 Hz, which rapidly increases to $\sim$20 Hz when the source
moves across the NB/FB vertex \citep{prhale1986,diva2000}. In recent
years two high frequency (or kHz) QPO were found in the Z sources
(\citealt{vaswzh1996,wihova1997,wihova1998,wimeva1998,jowiva1998,zhstsw1998},
see \citealt{va2000}, for a review). They are often observed
simultaneously, with a frequency difference of $\sim$300 Hz, and have
frequencies between 215 Hz and 1130 Hz, which increase from the HB to
the NB. In Sco X-1 the frequency difference was found to decrease
with increasing QPO frequency \citep{vawiho1997}. The frequency
difference in the other Z sources is both consistent with the
behavior seen in Sco X-1 and with being constant
\citep{wihova1997,jowiva1998,psmewi1998}. Three types of noise are
seen in the Z sources. They are the very low frequency noise (VLFN),
the low frequency noise (LFN), and the high frequency noise (HFN).
The VLFN and HFN are found on all branches, whereas the LFN is only
observed on the HB and NB. The VLFN, which is found at frequencies
below 1 Hz, can be described by a power law. The HFN and LFN are both
band limited components, with cutoff frequencies of, respectively,
10--100 Hz, and 2--10 Hz.

Many competing models have been proposed for the origin of the QPOs
and noise components. It is beyond the scope of this introduction to
mention these models in detail - most of them will be discussed in
Section \ref{gx17+2_discuss_sec}.

In this paper we present a study, based on data acquired with the
{\it Rossi X-ray Timing Explorer} (\xte), of the correlated spectral
and variability properties of the Z source GX 17+2. It is a
continuation of the work by \citet{wivaps1996,wihova1997}. The
current paper constitutes the first report on the very large
observing campaign of 1999, which more than doubled the total
coverage of the source. This campaign was undertaken with the express
purpose of investigating if a non-monotonic relation exists between
the frequency of the kHz QPOs and the HBO in GX 17+2. Section
\ref{17+2_obs_sec} deals with the observations and analysis. The
spectral results are presented in Section \ref{spectral_sec} and the
results for each power spectral component in Sections
\ref{timing_sec} and \ref{sec:res-khz}. A number of qualitatively new
results is found in our greatly expanded data set. In particular, we
find that when the upper kHz QPO frequency is below 1030 Hz, it
correlates with the HBO frequency, whereas above 1030 Hz they
anti-correlate. We also find that kHz QPO frequency difference is not
constant and that the Q values of the HBO and kHz QPOs are probably
not explained by life time broadening. These and other results are
compared to observations of other Z sources and discussed in terms of
current models in Section \ref{gx17+2_discuss_sec}. Finally, we argue
that changes in the spectral and variability properties of the Z
sources may not be not driven by changes in the mass accretion rate.

\section{Observations and Analysis}\label{17+2_obs_sec}

The data used for the analysis in this paper were all obtained with
the Proportional Counter Array (PCA, \citealt{jaswgi1996}) on board
\xte\ \citep{brrosw1993}. The PCA consists of five xenon-filled
proportional counter units (PCUs), each with an effective area of
$\sim$1250 cm$^2$ (at 10 keV). Although the five PCUs are in
principle identical they all have a slightly different energy
response. These responses change continuously due to slow processes
such as gas leakage and the aging of the electrodes. In addition, the
high voltage settings of the instruments are occasionally altered
(gain changes), resulting in rather more drastic changes in the
detector response. These changes have, at the time of writing, been
applied three times during the life time of \xte\, thereby defining
four gain epochs. Occasionally one or more PCUs are not operational.
They can be switched off by an internal safety mechanism, or by the
ground control crew, for reasons of detector preservation. Therefore,
the number of active detectors varies between the observations.

All our \xte/PCA observations of GX17+2 were done between 1997
February 2 and 2000 March 31. A log of the observations is given in
Table \ref{obs_tab}. We do not include the observations done in
February 1996, which were used by   \citet{wivaps1996}. The reasons
for this are the limited time resolution in the energy range of
interest and difficulties with scaling the \sz\ (see below) due to an
incomplete Z track.  Data taken during satellite slews and Earth
occultations were removed. The nine type I X-ray bursts that were
observed are the subject of a separate article by \citet{kuhova2001};
two of those were also studied in this paper. The total amount of
good data that remained was $\sim$600 ks.

Data were collected in several modes with different time and spectral
resolutions. Two of these modes, `Standard 1' and `Standard 2', were
always operational. In addition to these two modes, other modes were
active that varied between the observations. The properties of all
modes are given in Table \ref{gx17+2_modes_tab}.

The Standard 2 data were used to perform a broad-band spectral
analysis. The data were background subtracted, but no dead time
corrections were applied; these were in the order of 2--5\%.


For each 16 s data segment (i.e. the intrinsic resolution of the
Standard 2 mode) we defined two colors, which are the ratios of count
rates in two different energy bands, and an intensity, which is
simply the count rate in one energy band. The energy bands used for
the colors (soft and hard color) and intensity, are given in Table
\ref{colors_tab}. The lower energy boundaries for the soft color and
intensity were chosen relatively high in order to avoid the lowest
and least reliable energy channels. By plotting the two colors
against each other a color-color diagram (CD) was produced. A
hardness-intensity diagram (HID) was produced by plotting the hard
color against the intensity. To produce the CDs and HIDs we only used
data obtained with those PCUs that (for each gain epoch) were always
on. For gain epoch 3 these were PCUs 0, 1 and 2, and for gain epoch 4
PCUs 0 and 2. Due to the different numbers of detectors and the
differences in the detector settings we decided to produce the CDs
and HIDs separately for the two gain epochs. However, in choosing the
energy channels we tried to take channels whose energy boundaries
were as close as possible.

Due to the aging processes mentioned earlier, observations with the
same source spectrum that are made more than a few weeks apart end up
at a different location in the CD. To correct for this effect we
analyzed a number of \xte/PCA observations of the Crab pulsar/nebula
(which is assumed to have a constant spectrum, see also
\citealt{kuvaoo1994}) that were taken around the time of our GX 17+2
observations. For all Crab observations we produced the colors in the
same energy bands as we used for GX 17+2. We found that the observed
colors of the Crab indeed changed. For each observation we calculated
multiplicative scaling factors, for both colors, with respect to
those of the first Crab observation. We used these factors to scale
the colors of the GX 17+2 observations back to those of the first GX
17+2 observation. This procedure could only be applied to the epoch 3
observations, where, as expected, we found it to result in a narrower
track in the CD. No corrections were applied for the intensity.
Since no Crab observations were available for the March 2000
observations, no corrections could be applied to epoch 4 data. The shifts in that data set appeared to be small anyway.

\begin{figure}[t] \centerline{\includegraphics[width=15cm]{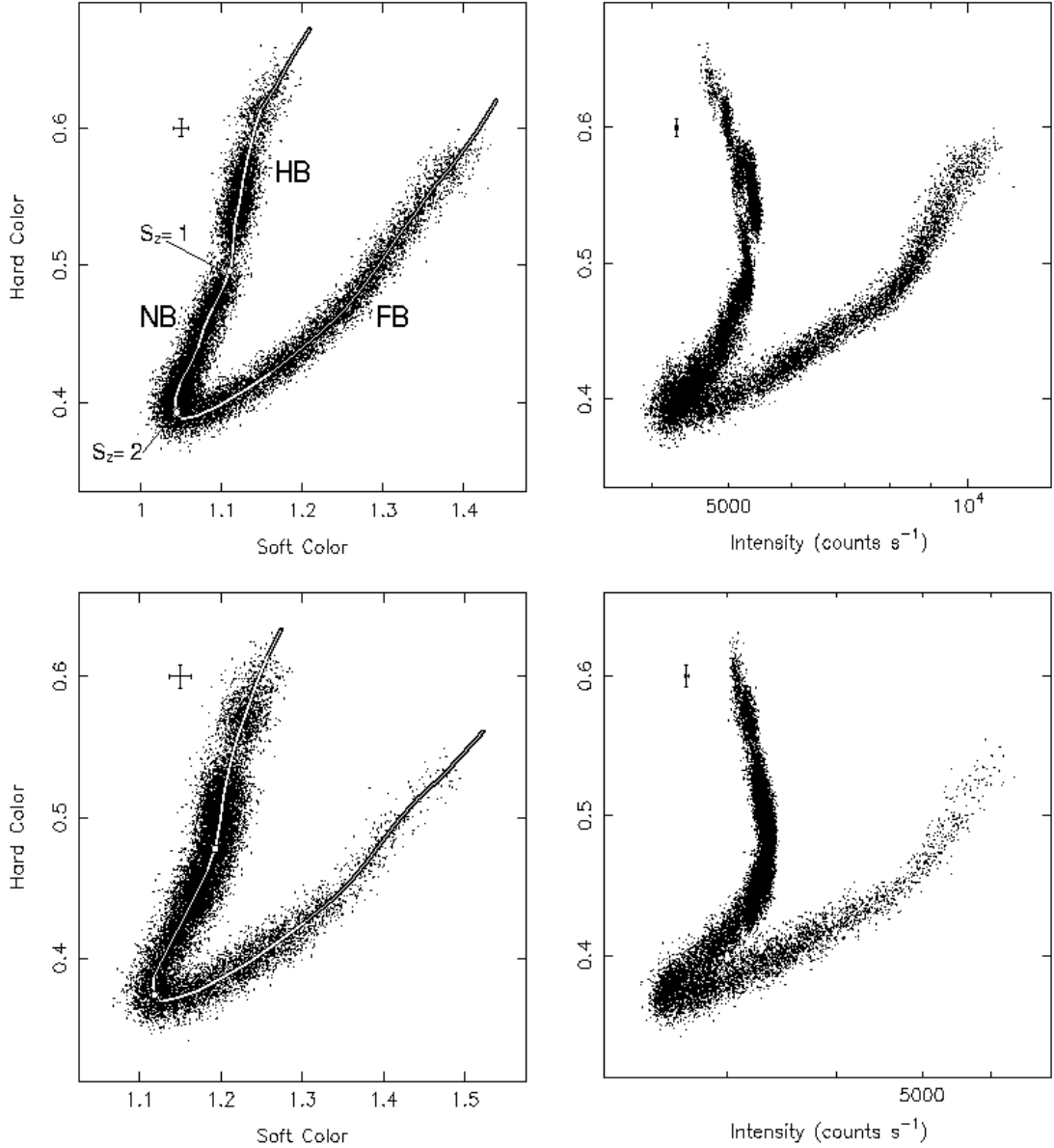}}
\caption{Color-color diagrams (left column) and
Hardness-Intensity diagrams (right column) for the epoch 3 (top) and
epoch 4 data (bottom). Each point represent a 16 s average. The
splines and vertices (white circles) that were used for the \sz\,
parameterization are shown, as are the typical error bars. See Table
\ref{colors_tab} for the energy bands used for the soft and hard
color and intensity.\label{cd-hid_fig}} \end{figure}

Our power spectral analysis was based on selecting observations as a
function of the position along the Z track in the CD. We used a
method that is based on the 'rank number' and '\sz' parameterization
methods introduced by \citet{havaeb1990} and \citet{hevawo1992},
that has been gradually refined in similar studies
\citep{kuvaoo1994,kuva1996,kuvaoo1997,wivaku1997,diva2000}. In this
method in its current form, all points in the CD are projected onto a
bicubic spline \citep[see][]{prteve1992} whose normal points are
placed by hand in  the middle of the Z track (see Figure
\ref{cd-hid_fig}). The HB/NB and NB/FB vertices are given the values
\sz=1 and \sz=2, respectively. The rest of the Z track is scaled
according to the length of the NB. Problems arise when sharp curves
are present in the track, in our study most notably around the NB/FB
vertex. Due to scatter, points which have FB properties end up on the
NB and vice versa. This is not a problem that is only intrinsic to
the \sz\, parameterization; it is a limitation that applies to all
selection methods based on colors. No observable parameter, apart
from the power spectra, has been identified that could be used to
distinguish NB and FB observations around the vertex better than the
X-ray colors.

In order to improve on the results of \citet{wihova1997} we wished to
combine the epoch 3 and epoch 4 data sets. Unfortunately, the tracks
traced out in the CDs of epoch 3 and 4 are not the same and two
different splines had to be used for the \sz\, parameterization. 
Since the normal points for these splines were drawn by hand, the
\sz\, scales for the epoch 3 and epoch 4 CDs were unlikely to be
exactly the same. Therefore, we first transformed the \sz\, scale of
epoch 4 to that of epoch 3.  To accomplish this, we measured the
frequency of either the HBO or NBO at several places along the Z
track of epoch 3, and determined the \sz\, interval corresponding to
the same frequency in the epoch 4 data. We found that $S_{z,epoch\,4}=(0.06\pm0.04)
+ (1.005\pm0.018)S_{z,epoch\,3}$, showing that, although the scales
are the same (within the errors), a small shift is present. We
subsequently scaled the $S_z$ values of epoch 4 using the above
expression. The above scaling method assumes that the HBO and NBO
frequencies are strongly related to \sz. Previous studies of Z
sources \citep[see
e.g.][]{kuvaoo1997,wihova1997,jovawi2000,diva2000}, as well as the
fact that only little scatter was present around the linear relation seem to confirm this.

Power spectra were created from the data in modes with high time
resolution ($\le2^{-13}\,s$; see Table \ref{gx17+2_modes_tab}) using
standard Fast Fourier Transform techniques \citep[see][and references
therein]{va1989}. The data were not background subtracted and no dead
time corrections were applied prior to the Fourier transformations.
We made power spectra in several energy bands, with several frequency
resolutions and Nyquist frequencies. We finally settled on
0.0625--4096 Hz power spectra in the 5.1--60 keV (epoch 3) and
5.8--60 keV (epoch 4) bands, since the QPOs were most significantly
detected in those bands (note that the source does actually not
contribute significantly to the power spectra above $\sim$40 keV,
however, our data modes did not allow us to discard data above this
energy). This choice of frequency range means that the properties of
the VLFN, which dominates the power spectrum below 0.1 Hz, could not
always be measured satisfactorily, but it does allow us to follow the
power spectral evolution on time scales down to 16 s. Note that the
power spectra of epoch 4 were produced in a slightly higher energy
band than those of epoch 3. Due to the gain changes and the limited
choice of energy channels this was the best possible match.

The power spectra were selected on the basis of $S_z$, as determined
from the CD. The $S_z$ selections usually had a width of 0.1 and did
not overlap - no power spectrum was represented in more than one
\sz-selection. Different widths were used in cases where the power
spectrum changed rapidly as a function of $S_z$ (narrower selections)
or when the powers were weak (wider selections). All the  power
spectra in a selection were averaged, and the resulting power
spectrum was rms normalized according to a procedure described in
\citet{va1995b}.  

The properties of the power spectra were quantified by fitting
functional forms to them. The low frequency (0.0625--256 Hz) and high
frequency (100--4096 Hz) parts of the power spectrum were fitted
separately. The high frequency part was fitted with one or two
Lorentzians (for the kHz QPO(s)) and with the function $P(\nu) = P_1 +
P_2 cos (2\pi\nu/\nu_N) + P_3 cos (4\pi\nu/\nu_N)$ for the dead time
modified Poisson level \citep{zh1995,zhjasw1995}; no separate term
was used for the contribution of the Very Large Events (VLE) count
rate since it was absorbed by this fit function. At low frequencies
the noise that was present  could not be fitted consistently, and
varying combinations of functionals had to be used:

\begin{itemize}
\item{\sz$<$0.0: A Lorentzian (LFN+VLFN)}
\item{\sz=0.0--0.1: A cut-off power law (LFN+VLFN)}
\item{\sz=0.1--1.4: A power law (VLFN) and a cut-off  power law (LFN)}
\item{\sz=1.4--1.6: A power law (VLFN) and a Lorentzian (LFN+NBO)}
\item{\sz=1.6--5.0: A power law (VLFN) and a Lorentzian (LFN)}
\item{\sz$>$5.0: A power law (VLFN)}
\end{itemize}

The expression for a power law is $P(\nu)\propto\nu^{-\alpha}$, that
for a cut-off power law is  $P(\nu)\propto \nu^{-\alpha}
e^{-\nu/\nu_{cut}}$ (where $\nu_{cut}$ is the cut-off frequency), and
for a Lorentzian $P(\nu) \propto 1/[(\nu-\nu_c)^2 + (FWHM/2)^2]$
(where $\nu_c$ is the centroid frequency and $FWHM$ is the
full-width-at-half-maximum). The HBO, its second harmonic and the
NBO/FBO were each fitted with a Lorentzian. Below the HBO, at about
half its frequency, a broad bump was present that was also fitted
with a Lorentzian, whose frequency was sometimes fixed to zero.  At
low frequencies the dead time modified Poisson level was fitted with
a constant. An example of a fit to a power spectrum (\sz=0.4--0.5),
displaying the contributing of the several components, is shown in
Figure \ref{pds_comp_fig}.

Note that changes in the fit function, such as using a cut-off power
law instead of a Lorentzian or adding an extra component, may lead to
minor changes in the values of other parameters. Errors on the fit
parameters were determined using $\Delta\chi^2=1$. Upper limits were
determined  by fixing some or all of the parameters of a component,
except the rms amplitude, to  values similar to those obtained in the
closest \sz\ selection where it was found to be significant,  leaving
all other fit parameters free, and  using $\Delta\chi^2=2.71$ (95\%
confidence).

A study of the energy dependence and time lag properties of the QPOs
and noise components will be presented elsewhere.

\section{Results}\label{results_sec}

\subsection{Broad-band spectral behavior}\label{spectral_sec} 

The CDs and HIDs of both epochs are shown in Figure \ref{cd-hid_fig}.
In both CDs the HB/NB vertex is not well defined, since the HB is
almost a continuation of the NB. This is mainly due to the relatively
high energies that we chose for our soft color (see Section
\ref{17+2_obs_sec}). At lower energies the turn-over is much clearer.
In Figure \ref{sz-count_fig} we show the count rate in several energy
bands as a function of $S_z$, for epoch 4. It shows that the HB/NB
vertex in the HID is entirely due to the count rates at low energies.
At high energies the HB is a perfect continuation of the NB, and no
vertex is present.

In the HID of epoch 3 the Z track appears to be segmented, which was
already noted by \citet{wihova1997}. Although no corrections for the
slow detector aging processes were applied to the intensity in this
HID, we note that a quick look showed that they would only have made
the segmentation more apparent. The shifts in the HID might be due to
the secular motion that has been observed in other Z sources. The
shifts, of up to $\sim$5\% in intensity, do not show up in the CD.
This is because  colors are ratios of intensities and are therefore
not very sensitive to overall intensity changes; exactly for this
reason CDs are preferable over HIDs for our purpose. Moreover, the
width of  the tracks in the CD is about 5\%, so changes smaller than
this are hard to observe.

The bottom panel of Figure \ref{hist-vz_fig} shows the distribution
of the time spent by the source in each part of the Z track. The
source spent 28\% of the time on the HB (\sz$\leq$1), 44.2\% on the
NB (1$<$\sz$\leq$2), and 27.8\% on the FB (\sz$>$2). The average
speed along the Z track as a function of \sz ($\langle V_z \rangle$)
is shown in the top panel of Figure \ref{hist-vz_fig}. The speed at a
given $S_z(i)$ is defined as $V_z(i)=|S_z(i+1) - S_z(i-1)|/32$
\citep[see also][]{wivaku1997}, where $i$ is used to number the
points in order of time.   As expected the $\langle V_z \rangle$
increases considerably when the source enters the FB, but
 also at the top of the HB ($S_z < 0.5$). Combined with
the small amount of time spent in the upper HB we can conclude that
the source reaches to \sz\ values this low only in brief, quick
dashes. 

For a more detailed analysis of the spectral changes as a function of
the position along the Z track in GX 17+2 we refer to \mbox{O' Brien} et al.
(in prep.).

\subsection{Power spectra}\label{timing_sec}

\subsubsection{Low Frequency QPOs and Noise Components}

Figure \ref{pds_fig} shows the low frequency part (0.0625--256 Hz) of
the power spectrum for nine different $S_z$ selections. The power
spectra shown in Figure \ref{pds_fig} are selected from epoch 3 and
epoch 4 and are therefore a combination of 5.1--60 keV and 5.8--60
keV data. The percentage of epoch 3 and epoch 4 data varies between
the \sz-selections. The contribution of epoch 3 data is highest at
the extremes of the \sz\ range, reaching 100\%, and gradually
decreases from both ends to $\sim$10\% around \sz=1.

Several components can be seen, which change in strength and shape as
a function of \sz. The different components are identified in Figure
\ref{pds_fig} (see also Figure \ref{pds_comp_fig}) and the \sz\
ranges in which they were detected are given in Table
\ref{sz-comp_tab}. The component identified as the sub-harmonic of
the HBO is rather broad and does not have the appearance of a QPO.
However, based on the frequency ratios (see below) it is referred to
as the HBO sub-harmonic. 

Some difficulties were experienced while fitting the low frequency
part of the power spectrum between \sz=1.4 and \sz=1.6, where two or
more components with similar frequencies were simultaneously present.
Above \sz=1.4 the NBO appeared, as a broad feature on top of the LFN.
We were not able to distinguish the two components and decided to fit
them together with a single Lorentzian. The fit values are not used
in figures and tables, since they do not represent any of the
individual components. Above \sz=1.6, the NBO and LFN could  be
distinguished more easily. The fit function used for the LFN, which
was underlying the NBO, was changed to a Lorentzian to be more
consistent with fits at higher \sz\ (fits with a cut-off power law
gave equally good $\chi^2_{red}$ at this \sz).

In the following sections the results for each of the components will
be presented.

\subsubsection{HBO}

The HBO was detected between \sz=$-$0.6 and \sz=2.1 and its second
harmonic between \sz=$-$0.6 and \sz=1.0. The second and third columns
of Figure \ref{hbo_fig} show their rms amplitudes, FWHM, and
frequencies as a function of \sz\,(see also Table \ref{low_tab}). The
frequency of the HBO increased from 21.3 Hz at \sz=$-$0.43 to 60.3 Hz
at \sz=1.45 and then decreased to 48.2 Hz at \sz=2.04. The second
harmonic of the HBO had a frequency that was on average
1.941$\pm$0.007 times that of the first harmonic, which is
significantly different from the expected value of 2. An  explanation
for this discrepancy is proposed in Section \ref{disc-hbo_sec}.
Between \sz=$-$0.6 and \sz=1.0 both the HBO and its second harmonic
decreased in strength, with the decrease of the second harmonic
occurring somewhat faster than that of the HBO. In that same \sz\
interval the FWHM of the HBO and its second harmonic were fairly
constant (showing a slight increase), although considerable scatter
was present around the average values (which were, respectively,
7.9$\pm$0.3 Hz and 14.4$\pm$0.6 Hz). The Q values of the HBO and its
harmonic are shown in Figure \ref{q-values_fig}. In the \sz\, range
where both were detected their Q values were consistent with each
other. When the second harmonic was not significantly detected
anymore (\sz$>$1.0) the rms amplitude and the FWHM of the HBO both
decreased from, respectively, 2.45$\pm$0.08\% and 9.7$\pm$7 Hz
(\sz=0.5--1.0) to 1.48$\pm$0.06\% and 5.1$\pm$0.4 Hz (\sz=1.0--1.5).
Between \sz=1.0 and \sz=1.5 the relation between the \sz\, and the
HBO frequency started to flatten. Above \sz=1.5 the frequency of the
HBO clearly dropped, initially quite abruptly and later on more
smoothly. This frequency drop coincided with an increase in the FWHM
to 15.7$\pm$1.5 (\sz=1.5--2.1); the rms amplitude showed a small
increase to 2.1$\pm$0.1\% (\sz=1.76), followed by a decrease to
1.2$\pm$0.2\% (\sz=2.04).

Underlying the HBO and its second harmonic, but with a central
frequency lower than that of the HBO, we found a broad feature that
was also fitted with a Lorentzian. It was significantly detected
between \sz=$-$0.6 and \sz=1.3.  The properties of the broad feature
are shown in the first column of Figure \ref{hbo_fig} (see also Table
\ref{low_tab}). Below \sz=0 the frequency of the Lorentzian was fixed
to zero. Between \sz=0.5 and \sz=1.5 the frequency of the broad
feature is on average 0.539$\pm$0.015 times that of the HBO. This
could mean that the broad feature is the sub-harmonic of the
HBO, certainly when one takes into account that the frequency of a
broad feature is rather sensitive to the shape of the continuum.
Whereas the rms amplitude and the frequency of the sub-harmonic both
change strongly with $S_z$ the FWHM remains more or less constant,
with an average value of 29.5$\pm$1.5 Hz.

\subsubsection{NBO and FBO}

The NBO and FBO were detected between \sz=1.6 and \sz=2.7.  The fit
results for the NBO/FBO are shown in Figure \ref{nbo_fig} (see also
Table \ref{low_tab}). Below \sz=2.1 the NBO (represented by the
filled circles in Figure \ref{nbo_fig}) had a fairly constant
frequency, with values between 6.3 Hz and 7.0 Hz. Its rms amplitude
increased from 1.7\% to 3.2\%, and the FWHM was approximately
$\sim$2.1 or $\sim$3.8. In the \sz=2.0--2.1 selection both the NBO
and FBO were present. This is likely an artifact of the \sz\
selection method, since a careful inspection of dynamical power
spectra showed no evidence for simultaneous presence of both QPOs. It
is interesting to note though, that in the \sz=2.0--2.1 selection the
frequency of the FBO was 2.0$\pm$0.1 times that of the NBO, which
could mean that the two NBO and FBO are harmonically related
(however, see below). The FBO increased in frequency from 13.9 Hz
(\sz=2.04) to 23.1 Hz (\sz=2.65), while its FWHM increased from
$\sim$5 Hz to $\sim$13 Hz. The rms amplitude of the FBO initially
continued the trend of the NBO rms amplitude; it increased from 2.8\%
(\sz=2.04) to 5.6\% (\sz=2.23), but then  decreased to 1.8\%
(\sz=2.65).

To study the transitions between the NBO and FBO more carefully, we
inspected all dynamical power spectra of observations with \sz\
values around 2. Although no clear transitions were found, mainly due
to the limited quality of the dynamical power spectra, we did in some
cases see QPOs with intermediate frequencies ($\sim$ 10 Hz),
suggesting that the frequency does not jump directly from $\sim$7 Hz
to $\sim$14 Hz. Fitting the power spectra in consecutive time
intervals (longer than 16 s), rather than inspecting dynamical power
spectra with the eye, will probably lead to more conclusive results,
but is beyond the scope of the current paper.  The time scales on
which the NBO/FBO frequency changed from $\sim$7 Hz to $\sim$14 Hz
and back were as short as a few tens of seconds.

We also studied the behavior of the NBO/FBO during two long type I
X-ray bursts. The first one started on 1998 November 18 at 08:51:26
UTC, the second one on 1999 Oct 10 at 09:10:47 UTC. Their exponential
decay times were, respectively, 189$\pm$2 s and 144$\pm$2 s
\citep{kuhova2001}. Both bursts occurred near the NB/FB vertex and in
the power spectra of their respective observations the NBO/FBO is
clearly detected. No other QPOs were detected in the power spectra of
these bursts. Figure \ref{burst-nbo_fig} shows the dynamical power
spectra of both bursts, together with their 2--60 keV light curves.
During the brightest part of the bursts the NBO/FBO seemed to
disappear. Apparently, the burst flux was not modulated at the
NBO/FBO frequencies with the same amplitude as the persistent flux.
To quantify this, we determined upper limits on the NBO/FBO strength
and compared those with the  values outside the bursts. The results
are shown in Table \ref{burst-nbo_tab}. Clearly, during the brightest
part of the bursts the rms amplitude of the NBO/FBO was significantly
weaker, not only as a fraction of the total flux, but also in
absolute terms (and hence as a fraction of the persistent flux, if
the persistent emission is assumed to continue during the bursts).
These measurements constitute the first determination of the effect of
X-ray bursts on the NBO/FBO. The fact that the bursts suppress the
QPO can have important consequences for our understanding of its
formation mechanism (see Section \ref{disc-nbo_sec}).

\subsubsection{Noise components: LFN and VLFN}

LFN was detected between \sz=$-$0.6 and \sz=5.0. As mentioned before,
the appearance of the NBO  around \sz=1.4, kept us from putting firm
constraints on the LFN parameters between \sz=1.4 and \sz=1.6. The
fit results for the LFN are shown in Figure \ref{lfn_fig} (see also
Table \ref{tab:noise}), in two separate columns: one for the fits
with a cut-off power law fit (\sz=0.0--1.4) and one for the fits with
a Lorentzian (\sz$<$0.0 and \sz=1.6--5.0). The strength of the LFN
was in both cases defined as the integrated power spectral density
between  1 Hz and 100 Hz. We note that below \sz=0.1 the VLFN was not
fitted separately from the LFN. An inspection of 1/256--4096 Hz power
spectra below \sz=0.1 showed that a weak power law component was
present at frequencies below 0.1 Hz. This component was probably
VLFN; its power in the 1--100 Hz range was much smaller than that of
the LFN, so, although some VLFN power was absorbed by the LFN, this
did not affect the LFN rms amplitudes significantly. The strength of
the LFN changed considerably as a function of \sz. It showed a narrow
peak between \sz=$-$0.6 and \sz=0.2 of $\sim$5\% rms. Between \sz=0.2
and \sz=1.7 it gradually decreased from $\sim$4.5\% rms to 
$\sim$2.5\% rms.  Another decrease was observed between \sz=2.7 and
\sz=5.0, from $\sim$3.8\% rms to $\sim$0.9\% rms. The behavior between
\sz=1.7 and \sz=2.7 was quite erratic, probably due to interactions
with the fit functions of NBO and FBO.

The centroid frequency of a Lorentzian and the cut-off frequency of a
cut-off power law cannot be directly compared. Since we wanted to see
how the typical frequency of the LFN evolved with \sz, we chose to
plot $\nu_{max}$, which is the maximum in a $\nu P(\nu)$ plot and the
frequency at which most of the power is concentrated per logarithmic
frequency interval (Belloni, Psaltis, \& van der Klis 2001, in
prep.). For a Lorentzian  $\nu_{max}$ is $(\nu_c^2 +
(FWHM/2)^2)^{1/2}$ and for a cut-off power law $\nu_{max}$ is
$(1-\alpha)\nu_{cut}$ (see Section \ref{17+2_obs_sec} for analytical
expressions for a Lorentzian and cut-off power law). As can be seen
from Figure \ref{lfn_fig}, between \sz=$-$0.6 and \sz=1.7 
$\nu_{max}$ smoothly increased from $\sim$1.2 Hz  to $\sim$14.7 Hz. 
Above \sz=1.7 the errors on $\nu_{max}$ were larger and the behavior
was less clear. Between \sz=1.7 and \sz=2.3 $\nu_{max}$ decreased to
$\sim$6.5 Hz, and above \sz=2.3 it increased again to $\sim$15 Hz. We
tested whether the change of the fit functions at \sz=0.0 affected 
the values for $\nu_{max}$ and the  other power spectral parameters,
by swapping the fit functions used below and above \sz=0.0; no
significant changes were found.

VLFN was detected over almost the whole \sz\ range. Although we only
started fitting the VLFN separately from the LFN above \sz=0.1 (in
the 16 s power spectra), VLFN was present at frequencies below 0.1 Hz
in the  256 s power spectra below \sz=0.1. In the 16 s power spectra
it was only significantly detected above \sz=0.3. The fit results for
the VLFN are shown in Figure \ref{vlfn_fig} (see also Table
\ref{tab:noise}). Between \sz=0.3 and \sz=1.4 the VLFN strength
decreased from $\sim$0.9\% rms to $\sim$0.4\% rms. Above \sz=1.4 its
strength increased, to a peak value of $\sim$1.3\% rms at the NB/FB
vertex. On the lower FB the strength decreased again, to a value of
$\sim$0.5\% rms, and above \sz=3.0 it increased to $\sim$1.6\% rms at
the top of the FB. The index of the VLFN slowly increased from
$\sim$0.6 to $\sim$1.0 between \sz=0.1 and \sz=2.4, with a small peak
between \sz=1.6 and \sz=1.9, where the index had a value of
$\sim$1.5. Between \sz=2.4 and \sz=3.2 the VLFN was much steeper. The
indices were not well constrained and had values between 1.7 and 4.0.
Above \sz=3.2, where the error bars are much smaller, the index
slowly decreased from 2.1 to 1.8. 

Our fits did not require an additional component to fit possible HFN;
no such component was found. Possible explanations for this are
discussed in Section \ref{disc:noise}.

\subsection{High Frequency QPOs}\label{sec:res-khz}

Both the lower and upper kHz QPO were clearly detected; the lower kHz
QPO  between \sz=0.5 and \sz=1.5, the upper kHz QPO between \sz=--0.3
and \sz=1.7. The results can be found in Table \ref{khz_tab} and are
shown in Figure \ref{khz_fig}. The QPO frequencies showed a clear
increase with \sz, although both relations flattened at their low
frequency ends. The FWHM of the lower kHz QPO was consistent with
being constant at $\sim$100 Hz, whereas the rms amplitude showed a
peak near \sz=1.05 with a value of 3.6$\pm$0.3. The rms amplitude and
FWHM of the upper kHz QPO both decreased with \sz.

Figure \ref{khz-diff_fig}a shows the frequency difference of the two
kHz QPOs  as a function of the frequency of the upper kHz QPO. Some
fits were made to the data; they are also shown in Figure
\ref{khz-diff_fig}a. The best fit to the frequency difference with a
constant gave a value of 282$\pm$4 Hz. The $\chi^2/d.o.f.$ for this
fit was 18.6/9, which means that at a 97\% confidence level the
frequency difference was not constant. This is the first time that
this is observed for GX 17+2. Fits with first and second order
polynomials (taking into account errors in both coordinates) resulted
in,  respectively, $\chi^2/d.o.f.$=15.0/8 and $\chi^2/d.o.f.$=3.7/7.
Although the latter two fits show that with 99.76 \% confidence
(3.2$\sigma$) a decrease towards higher frequency is not monotonic,
it is not clear either whether the decrease towards lower frequencies
is significant or not. This would be the first time that such a
decrease towards lower frequencies is observed in any kHz QPO source
with a non-constant frequency difference. To test this, two fits with
a broken line (not shown) were performed, where in one case the slope
of the low frequency part was fixed to zero. They resulted in
$\chi^2/d.o.f.$=6.28/7 (slope fixed) and $\chi^2/d.o.f.$=2.52/6
(slope free), which suggests that the frequency decrease towards
lower frequency is only significant at a 97.6\% confidence level
(2.4$\sigma$).

In Figure \ref{khz-diff_fig}b we show additional fits to the data
with theoretical curves for the radial epicyclic frequency vs.\
azimuthal frequency, which in the relativistic precession models are
associated with the frequency difference and the upper kHz QPO
frequency. The solid line is the best fit for circular orbtis
\citep{stvi1998,mala2001}, which  yields a $\chi^2_{red}$ of 7.3
(d.o.f = 9) for a mass of 2.07 M$_\odot$; the dashed line is the best
fit for eccentric orbits \citep{mala2001}, yielding a $\chi^2_{red}$
of 1.14 (d.o.f = 8) for a mass of 1.86 M$_\odot$. Both curves were
calculated for non-rotating stars; rotation of the neutron star gives
only insignificant improvements. For comparison we also plotted the
data for Sco X-1 from \citet{vawiho1997}. Clearly, both fits (especially the solid line) do not fit the decrease of
the frequency difference towards lower frequencies, which occurs in a
frequency range for the upper kHz QPO that was not observed in Sco
X-1, very well. 

The Q values (frequency/FWHM) of the two kHz QPOs were consistent
with each other and with that of the HBO and its harmonic (see Figure
\ref{q-values_fig}); between \sz=1.0 and \sz=1.5 the Q values of the
HBO and kHz QPOs increased from $\sim$5 to $\sim$10. This unexpected
finding may provide a key to understanding the formation mechanism of
the three QPOs -- this will be further discussed in Section
\ref{q-values_sec}.

When comparing Figures \ref{khz_fig} and \ref{hbo_fig} one can see
that above \sz=1.5, where the HBO frequency starts to decrease, the
upper kHz QPO frequency still increases. In Figure \ref{khz-hbo_fig}
we plot the frequencies of both QPOs against each other. For values
of the upper kHz frequency lower than 1030 Hz the frequencies are
well correlated, but above there is a clear anti-correlation. The
solid line is the best power law fit to the points below 1000 Hz. The
power law index is 2.08$\pm$0.07. This is the first time in any atoll
or Z source showing kHz QPOs that an anti-correlation is observed
between the frequencies of the low and high frequency QPOs (see
Section \ref{disc-hbo_sec} for a discussion).

\section{Discussion}\label{gx17+2_discuss_sec}

We performed a detailed study of the low and high frequency power
spectral features of the neutron star low-mass X-ray binary and Z
source GX 17+2. As was found in previous studies of GX 17+2 and other
Z sources, the properties of most power spectral features correlated
well with the position of the source along the Z track in the X-ray
color-color diagram, suggesting that the spectral and variability
changes are caused by one and the same parameter. Some new results
were found, the most interesting being  the fact that the frequency
separation of the kHz QPOs was not constant and probably not
monotonic (Fig.\ \ref{khz-diff_fig}), that their Q values were
consistent with those of the HBO and its harmonic (Fig.\
\ref{q-values_fig}), and that the frequency of the upper kHz QPO was
anti-correlated with that of the HBO when the latter started to
decline on the NB (Fig.\ \ref{khz-hbo_fig}). We also found a
sub-harmonic of the HBO and showed that the absolute amplitude of the
NBO/FBO is suppressed during type I X-ray bursts (Fig.
\ref{burst-nbo_fig}). In the remainder of this section these and
other results will be discussed in more detail.  For a more detailed
discussion, including a comparison of our results with those obtained
from with other satellites (like \exo\, and \gin ) and of other Z
sources, we refer to \citet{ho2001}.

\subsection{Timing behavior}\label{disc_timing_sec}

\subsubsection{HBO and kHz QPO models}

In our discussion of the HBO and kHz QPO results, we will only
consider models that try to explain the behavior of both types of
QPOs. Currently these models are the relativistic precession model,
the sonic point beat frequency model and the two-oscillator model. In
the latter model \citep[see][]{osti1999} an angle between the
magnetospheric equator and the disc plane is defined, which depends
on the frequencies of the HBO and kHz QPOs and is supposed to be
constant. However, \citet{jovaho2001} show, by using our GX 17+2
results and theirs of GX 5-1, that this is definitely not the case.
For this reason, we will only consider the first two models. In the
relativistic precession model (RPM,
\citealt{stvi1998,stvi1999,most1999,stvimo1999}) three frequencies
are predicted; they correspond to the fundamental frequencies of
motion of a test particle in the vicinity of a compact object and are
identified with the HBO and the two kHz QPOs. The lowest predicted
frequency, which should explain the HBO, is a factor 2 lower than the
low frequency QPOs in the atoll sources and the sub-harmonic of the
HBO in the Z sources. Hence, a way to produce second harmonics is
required. Note that no actual mechanism is proposed for the
production of QPOs themselves. However, in a recent model by
\citet{psno2001} the accretion disk acts as low band-pass filter with
resonances close to the three frequencies predicted in the RPM. It is
unclear to what extent the details of the RPM (i.e. the turnover in
the relation between HBO and upper kHz frequency) carry over to the
\citet{psno2001} description. We also note that recent investigations
of the RPM by \citet{mala2001} have cast some doubt on the
frequencies predicted by the RPM  and relations between them. In the
sonic point beat frequency model
\citep[SPBFM,][]{milaps1998,lami2001} the HBO is explained by the
original beat frequency model of \citet{alsh1985} and
\citet{lashal1985}. The observed kHz frequencies are close to the
orbital frequency at the sonic point of this disk and the beat of
this frequency with neutron star spin frequency. The model requires
the presence of a magnetosphere and a solid surface, and can
therefore not explain the low and high frequency QPO in the black
hole candidates.

\subsubsection{Horizontal Branch Oscillations}\label{disc-hbo_sec}

The decrease of the HBO frequency on the NB in GX 17+2 was first
observed in \xte\ data that were not used in our analysis
\citep{wivaps1996}. There were interesting features in the HBO
properties which were reproduced in that data and ours. In both cases
the \sz\, dependence of the HBO frequency was not symmetric around
its peak value; below, the frequency slowly flattened off to its
maximum value, but above it showed a fast decrease by a few Hz that
was followed by a more or less linear decrease with \sz. Also,
simultaneously with the frequency drop, the FWHM increased by a
factor of more than two. Possible explanations for the HBO frequency
decrease are discussed in Section \ref{hbo-khz_disc}.

In all Z sources where an harmonic of the HBO has been observed, the
frequency ratio of the harmonic and the fundamental is generally less
than two. This can be explained, at least in GX 17+2, by looking at
the \sz\, dependence of the rms amplitudes of both components. In a
certain \sz\, selection both frequencies tend to be dominated by
those with the higher rms amplitudes, and hence by lower \sz\, and
lower frequencies; this effect will be stronger for the harmonic than
for the fundamental, since the \sz\ dependence of its rms amplitude
is steeper  (see Figure \ref{hbo_fig}). As a result the frequency
ratio will end up to be less than two. To test this hypothesis we
calculated the weighted averages of two harmonically related linear
functions, representing the HBO frequency-\sz relation between
\sz=0.0 and \sz=1.0. As weight we used linear functions that were fit
to the HBO rms-\sz\ relation between \sz=0.0 and \sz=1.0. The ratio
of the weighted frequency averages was 1.93, which is very close to
the observed value of 1.94. 

Peaked features with frequencies intermediate to those of the LFN and
the HBO have been reported for GX 340+0 \citep{jovawi2000}, Sco X-1
\citep{vawiho1997,wiva1999a}, GX 5-1 \citep{kuvaoo1994,jovaho2001},
and now also for GX 17+2. Based on the average frequency ratio of
this sub-HBO component and the HBO in GX 17+2, 0.539$\pm$0.015, we
suggest that, similar to what was suggested for GX 340+0, it may be
the sub-harmonic of the HBO.   A comparative study by
\citet{wiva1999a} of low frequency (HBO-like) QPOs and band limited
noise components in the neutron star Z and atoll sources and black
holes revealed strong correlations between the typical frequencies of
those components. When plotting the frequency of the low frequency
QPO versus the break frequency of the band limited noise they found
that the branch traced out by the Z sources, although lying parallel
to that of the atoll source and black holes, was slightly offset. It
is interesting to see that this discrepancy between the Z sources on
the one hand and the atoll sources and black holes on the other
disappears \citep{jovawi2000}, at least for GX 17+2, Sco X-1 and GX
340+0, when one uses the frequencies of the sub-HBO component,
instead of the HBO frequency \citep[see also][]{wiva1999a}. If the low
frequency QPO in the atoll sources and the sub-HBO component of the
three Z sources would indeed be the same QPO, this would mean that
the(se) Z sources do not anymore follow the relation between the low
frequency (HBO-like) QPO and the lower kHz QPO as defined by atoll
and black hole systems \citep{psbeva1999}. It is interesting to see
that the three Z sources would then end up near or slightly above a
second relation that is traced out by atoll sources like  4U 1728-34
and 4U 1608--52. At the moment, however, it seems that the relation
found by \citet{wiva1999a} and the main relation of 
\citet{psbeva1999} are mutually exclusive for the two types of
sources. Note that \citet{wiva1999a} found that if in their plot for
Sco X-1 the frequencies of the sub-HBO component and the HBO were
used, instead of the LFN and HBO frequencies, the discrepancy between
Sco X-1 and the non-Z sources also disappeared. This reinterpretation
has the advantage of preserving the consistency with the
\citet{psbeva1999} relation as well.

The dominance of the even harmonics over the odd harmonics in the low
frequency (HBO-like) QPOs, as suggested by the presence of a
sub-harmonic of the HBO, is not only observed in the Z sources, but
also a common feature in BHCs such as GS 1124--68 \citep{bevale1997}
and XTE J1550--564 \citep{howiva2001}. It suggests that a two-fold
symmetry is present in the production mechanism of these low
frequency QPOs. For the SPBFM one could think of an asymmetry in the
magnetic field, which results in different areas for the polar caps
\citep[see e.g.][]{kuvaoo1997}. In the RPM a two-fold symmetry is
introduced by the two points at which the slightly inclined orbits of
test particles cross the plane of the accretion disk.

\subsubsection{High frequency QPOs}

The kHz QPOs in GX 17+2 were discovered with \xte\,
\citep{vahowi1997,wihova1997}. In our data the lower kHz QPO was
found between 517 Hz and 794 Hz (\sz=0.5--1.5) and the upper kHz QPO
between 618 Hz and 1087 Hz (\sz=-0.3--1.7), similar to what was
reported by  \citet{wihova1997}. Our findings of the kHz QPOs in GX
17+2 are compatible with the general properties of kHz in other Z
sources \citep{va2000}.

An important new result is that  the frequency difference is no
longer consistent (at a 97\% confidence level) with being constant
(see Figure \ref{khz-diff_fig}). The frequency separation reached a
maximum  value of 308 Hz, when the upper peak frequency was about 915
Hz, and fell off in both directions to values of 263 Hz and 239 Hz,
when the frequencies of the upper peak were, respectively, 780 Hz and
1033 Hz. Note that the decrease in both directions, when fitted with
a parabola, resulted in significantly (3.1 $\sigma$) better fits than
that with a constant. The decrease in the frequency separation at the
low frequency side of the upper kHz QPO, although itself only
significant at a 2.3$\sigma$ level, has not been seen before in any
kHz QPO source. In the other LMXBs that show an non-constant
frequency separation, Sco X-1 \citep{vawiho1997}, 4U 1608--52
\citep{mevava1998,mevawi1998}, 4U 1735--44 \citep{fovava1998}, and 4U
1728--34 \citep{meva1999}, the frequency separation  was not measured
at upper peak frequencies as low as in GX 17+2. Although both the
SPBFM  and the RPM  are capable of explaining the observed decrease
in frequency difference towards higher upper peak frequencies, they
have severe problems with a  decrease towards lower frequencies. Such
a decrease is not possible in the SPBFM, and, although it is in
principle predicted by the RPM, that model only yields reasonale fits to our data when one allows for highly eccentric orbits (see Figure \ref{khz-diff_fig}b).

\subsubsection{HBO-kHz QPO frequency relation}\label{hbo-khz_disc}

The principal aim of our observing campaign in 1999 was to check on
the relation between HBO frequency (whose increase was known to turn
into a decrease towards higher \sz) and upper kHz QPO frequency. 
Clearly, a decrease, or even flattening, of the upper kHz QPO
frequency, is not observed to accompany the decrease in HBO frequency
(see Figures \ref{hbo_fig}, \ref{khz_fig}, and \ref{khz-hbo_fig}). In
a pure RPM, this can not be explained. However, if classical
precession due to oblateness of the neutron star also plays a role,
such an explanation is possible. For certain combinations of neutron
star parameters, the prediction is that above a certain value of the
upper kHz QPO frequency, the frequency of the HBO in fact starts
to decrease \citep[see][]{most1999}, due to the increasing importance
of the classical precession. In GX 17+2 this turn-over of the HBO
frequency occurs at a value of $\sim$1 kHz for the upper kHz QPO with
a peak frequency of $\sim$60 Hz for the HBO. Judging from Figure 2 in
\citet{most1999}, the relation between the HBO and upper kHz QPO
frequencies in GX 17+2 would require a high neutron star spin
frequency ($\sim$500--700 Hz), a neutron star mass of more than 2
$M_\odot$, and a hard equation of state (assuming the observed HBO
frequency is four times the predicted frequency). In the RPM, well
below the turn-over, the HBO frequency is expected to scale
quadratically with the upper kHz QPO frequency.  The best power law
fit to the data below 1000 Hz yielded an index of 2.08$\pm$0.07 (see
Figure \ref{khz-hbo_fig}), which is consistent with a quadratic
relation.  

In the SPBFM a turnover in the HBO-upper kHz QPO relation is not
predicted, but as the two frequencies do not directly depend on each
other, the model is flexible enough to account for it; in case the
turnover in HBO frequency can be explained by the two flow model put
forward by \citet{wivaps1996}, the steady increase of the upper kHz
QPO frequency could in principle be explained by assuming that most of
the radial flow falls back to the disk and rejoins the disk flow
before the sonic point is reached, so that \mdot\ at that location,
and hence the upper kHz QPO frequency keep increasing. However, since the two radii at which the different QPOs are generated are very close, this seems a rather unlikely option.

\subsubsection{Q values of the HBO and kHz QPOs}\label{q-values_sec}

The Q values of the kHz QPOs are remarkably similar to each other,
and to those of the HBO and its harmonic (see Figure
\ref{q-values_fig}). They also show an almost simultaneous increase 
above \sz=1, when the HBO harmonic is no longer detected. If the QPOs
were artificially broadened by our selection method, averaging
together peaks with different centroid frequencies, one would expect
correlations of the FWHM (and Q value) with the width of the \sz\,
selection and with $d\nu/dS_z$. Such correlations, if present at all,
are far too weak to explain the observed similarities. 

The FWHM of a QPO can be caused by several mechanisms; intrinsic
frequency variations, phase jumps, the simultaneous presence of
several frequencies, and a finite lifetime of the signal will all
cause a signal to appear as a broadened peak in a power spectrum. The
data quality did not allow us to test the phase jump option. Lifetime
broadening can probably not explain our results. In the SPBFM there
is no reason why the lifetimes of the low and high frequency
oscillations, which are determined by different processes,  would
lead to similar Q values. In the RPM, where all frequencies are
generated by a single blob of matter, the lifetime for all variations
should to first order be the same. However, in that case the Q values
of the kHz QPOs should be much higher (by a factor 10--20) than those
of the HBO, since more kHz cycles fit in a blob's lifetime. Of
course, it is possible that the lifetime of the oscillations is not
determined by the lifetime of the blob, but rather by a damping
factor. Assuming that these factors are not the same for the
different QPOs, the similarity of the Q values could in principle be
explained. However, there is no obvious reason why the (independent)
damping factors would be fine tuned in a such a way that they lead to
similar Q values.

Explaining the Q values by broadening due to the presence of several
frequencies does not work either. Below the turnover in the 
HBO-upper kHz QPO frequency relation, the frequency of the HBO scales
quadratically with the frequency of the upper kHz QPO. A given range
of upper kHz frequencies would therefore lead to a Q value for the
HBO that is half the Q value for the upper kHz QPO. This is not
consistent with what we find.

Frequency modulation could in principle explain the similar Q values,
but only if the modulation is caused by a different parameter than
the one that determines the frequency changes along the Z track.
Frequency modulation only produces similar Q values if the QPO
frequencies vary with the modulating parameter by the same factor,
i.e.\ proportionally to each other. The observed quadratic relation
between the HBO frequency and that of the upper kHz QPO as well as
the subsequent turnover, shows that this is not the case for the
parameter that causes the motion along the Z track. However, assuming
that the modulating parameter is not responsible for motion along the
Z (see Section \ref{disc-spec}), we discuss the possibility of
frequency modulation in a bit more detail. For broadening by
frequency modulation to be observable in our power spectra, the time
scale of these variations should be longer than that of the HBO.
Additionally, the strength of those variations should anti-correlate
with the Q values of the QPO. Assuming that the parameter that causes
the frequency modulations also causes some changes in the count rate
we could try to identify it in the power spectrum. The only
reasonably strong components that we see in our power spectra and
whose typical frequencies fulfill the above frequency requirement
($\nu<10$ Hz) are the VLFN, the LFN and the NBO/FBO. The VLFN might
be associated with the motion along the Z (see Section
\ref{disc:noise}); disentangling that effect from a possible
additional one on QPO frequencies is beyond the scope of this work.
The strength of the LFN, which is of interest below \sz=1.5, 
anti-correlates with the Q values below \sz=1.5. Above \sz=1.5 the Q
values decrease again to a value between 2 and 5. In that case the
broadening might be due to the appearance of the NBO/FBO and the
increase in the LFN strength. It is interesting to note that in Sco
X-1 it has already been found that the kHz QPO properties are clearly
modulated by the NBO \citep{yuvajo2001}.

\subsubsection{Normal and flaring branch
oscillations}\label{disc-nbo_sec}

For the first time we were able to study the properties of the
NBO/FBO during type I X-ray bursts. We found that the absolute rms
amplitude significantly decreased when the radiation from the neutron
star surface increased. This is clearly different from the behavior
of the $\sim$1 Hz QPOs in the dipping LMXBs 4U 1323--62
\citep{jovawi1999}, EXO 0748--676 \citep{hojowi1999}, and  4U
1746--37 \citep{jovaho2000}, whose absolute rms amplitudes increased
during type I X-ray bursts. It is thought that these QPOs are not
related to the NBO/FBO in the Z sources. The behavior of the NBO/FBO
in GX 17+2 shows that NBO/FBO mechanism is very sensitive to the
radiation field, thereby lending strong support for radiation
feedback mechanisms (see below). In the framework of such mechanisms
it is easy to understand that the NBO/FBO disappears during bursts:
the NBO/FBO is only observed in a narrow \sz\, range, suggesting that
it requires a delicate balance of certain parameters, among which the
radiation field. The observations during the burst tails indicate
that the NBO/FBO mechanism does not switch on instantly but rather
becomes stronger in a more gradual way. Spectral fits during the
bursts \citep{kuhova2001} suggest that the spectral properties of the
accretion flow itself are not affected much by the increase in
radiation.

The frequency of the NBO/FBO changed from being more or less constant
below \sz=2 to being strongly correlated with \sz, above \sz=2. In
our data set we found the NBO between \sz=1.6 and \sz=2.1, although
it may already have been present as a very broad feature as early as
\sz=1.4. Between \sz=2.0 and \sz=2.1, the NBO (6.86 Hz) was detected
simultaneously with the FBO (13.9 Hz). Above \sz=2.1 the FBO
frequency increased, to $\sim$23 Hz at \sz=2.65. Note that the
simultaneous presence of the NBO and FBO between \sz=2.0 and \sz=2.1
is likely an artifact of the \sz\ selection method. Inspection of
dynamical power spectra showed that they were never present at the
same time. It is interesting though, that in that \sz\ selection
their frequencies differed by a factor of 2, which is suggestive of
mode switching of the NBO/FBO. However, there is evidence for QPOs
with intermediate frequencies, which undermines that idea.  The only
way to see whether the frequencies  really  gradually transform into
each other is to study the  power spectral changes as a function of
time. However, the quality of the GX 17+2 data is not sufficient to
perform such a study.

The presence of the NBO/FBO in the Z sources has often been related
to the fact that these sources accrete at near Eddington mass
accretion rates. It is thought that at these high mass accretion
rates a significant fraction of the accretion flow is in the form of
a thick and perhaps spherical flow. Also, the effects of radiation
pressure are thought to play an important role in this regime. The
model for the NBO by \citet{folami1989} is based on a radiation
feedback mechanism in a spherical accretion flow. 
Another model for the NBO was proposed by \citet{alhash1992}, in
which the NBO frequency is basically that of sound waves in a
thickened accretion disk. This model does not explain how the NBO
changes into the FBO. A major problem for both models could be the
recently discovered NBO/FBO-like QPOs in the two atoll sources 4U
1820--30 \citep{wivari1999} and XTE J1806--246 \citep{wiva1999b}.
Certainly in the first source the luminosity is believed to be only
$\sim$20--40 \% of the Eddington luminosity, while near-Eddington
luminosities are required in both models.

\subsubsection{Noise Components - LFN, VLFN}\label{disc:noise}

The lack of HFN in our data, compared to, e.g., in the \exo\
\citep{kuvaoo1997} data can be explained by instrumental effects and
different approaches to fitting the data. A study of {\it EXOSAT}
data of several sources by \citet{beva1994} showed that a component
with an rms amplitude of $\sim$3\% is always present in the frequency
range where the HFN was often found in the \exo\ data. When HFN was
found at lower frequencies, it was, based on strength and frequency,
most likely the component that we identified as the sub-harmonic of
the HBO.

Past measurements of the energy spectrum of the VLFN suggest that it
is caused by motion of the source along the Z
\citep{va1986,va1991,leluta1992}. The strength of the VLFN (Fig.
\ref{vlfn_fig}) and  the speed along the Z track ($\langle V_z
\rangle$, see Figure \ref{hist-vz_fig}) should therefore have a
similar dependence on \sz. Our results are inconclusive; although
some correlations are observed (the increase above \sz=3, and the
steep increase around \sz=2) there are also some differences. Most
notable is the dip in the VLFN strength around \sz=3, which is not
observed as a decrease in $\langle V_z \rangle$. However, it is
possible that due to the steepness of the VLFN around \sz=2.5 most of
its power was outside the range in which we determined the strength. 

The LFN followed the behavior of the HBO, both in strength and
frequency, until the latter disappeared ($S_z=2.1$). Between \sz=2.1
and \sz=2.8 the behavior is rather complex and probably considerably
influenced by the presence of the NBO/FBO.  Above \sz=3.0 the
behavior was rather clean, with a decrease in strength and an
increase in frequency. It is not clear, however, whether the LFN on
the FB is the same as that on the HB and NB. Unfortunately no HBO was
observed above \sz=2.1 to compare its behavior with.    Van der Klis
et al.\ (1997) \nocite{vawiho1997} suggested that in Sco X-1 the
NBO/FBO emerged from the LFN. This seems not to be the case in GX
17+2, since the rms amplitude of the LFN already started to decrease
well before the NBO appeared. Moreover, $\nu_{max}$ shows a gradual
increase between \sz=-0.6 and \sz=1.4 to values well above the NBO
frequency. A comparative study by \cite{wiva1999a} strongly
suggests that LFN  has a similar origin as the  band-limited noise
component in atoll sources, BHCs, and the millisecond X-ray pulsar
SAX 1808.4-3658. They suggest that this noise component is produced
in the accretion disk, and does not require the presence of
magnetosphere or solid surface.

\subsection{Broad-band Spectral Behavior}\label{disc-spec}

\begin{figure}[t]
\centerline{\includegraphics[width=10cm]{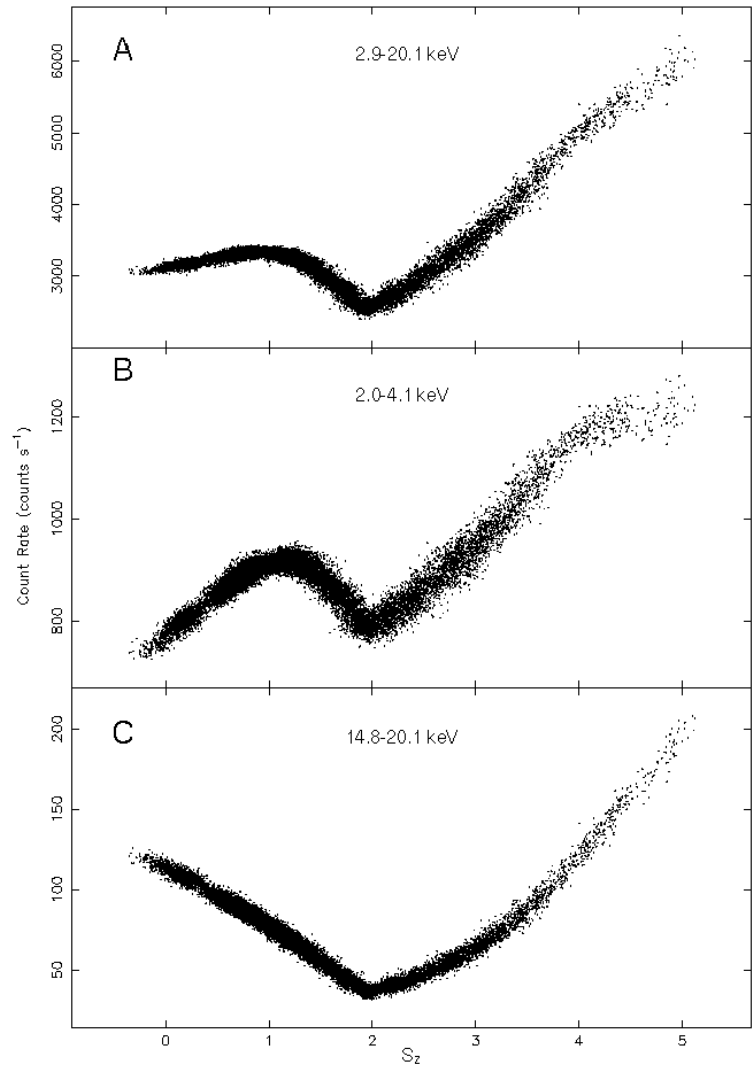}}
\caption{The count rate in three  energy bands as
a function of \sz\, (epoch 4 data only). The HB/NB vertex in the
2.9--20.1 keV band (a) is caused by the contribution from the low
energies (b). This vertex does not show up at high energies (c).
\label{sz-count_fig}} \end{figure}

\begin{figure}[t]
\centerline{\includegraphics[width=10cm]{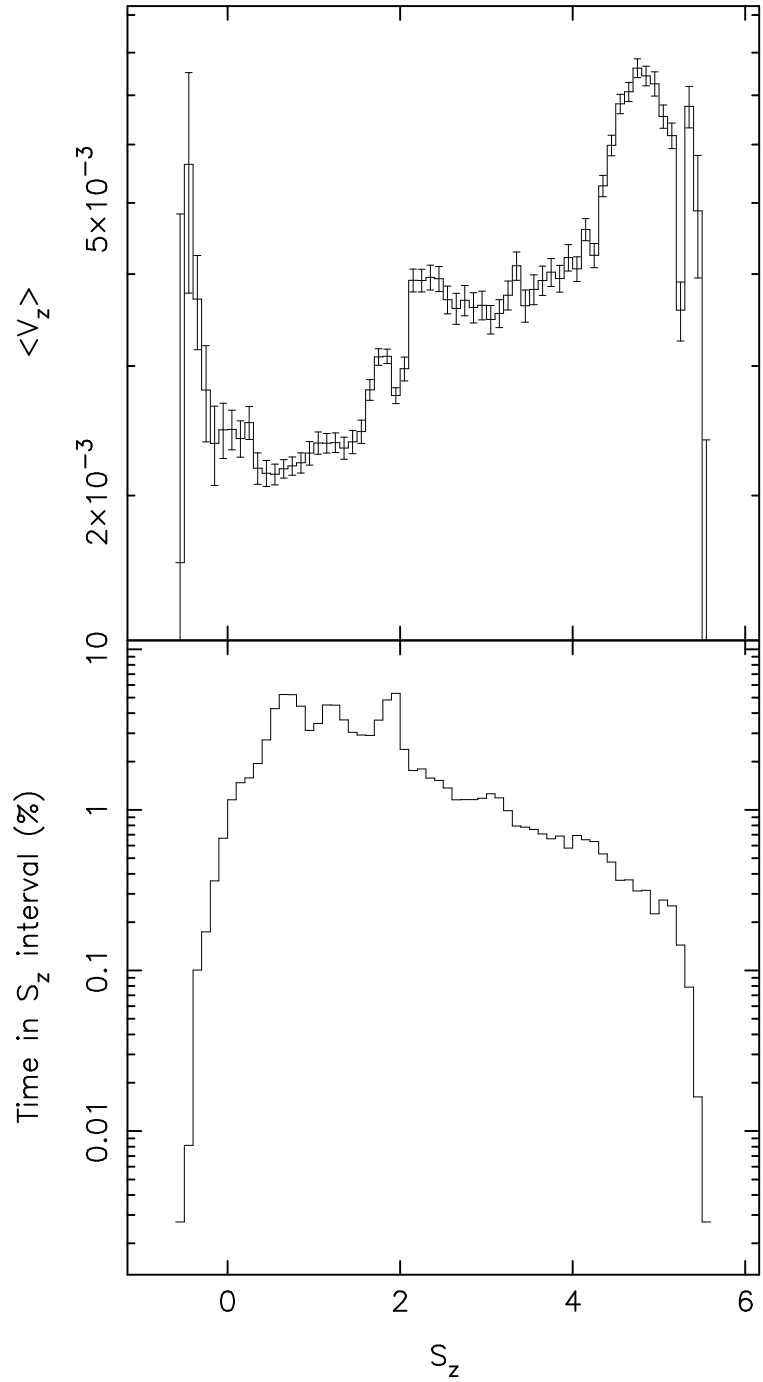}}
\caption{{\it Top}: The average speed along the Z track ($\langle V_z
\rangle$). {\it Bottom}: the percentage of the time spent in each
\sz-interval as a function of \sz.\label{hist-vz_fig}} \end{figure}

\begin{figure}[t] \centerline{\includegraphics[width=10cm]{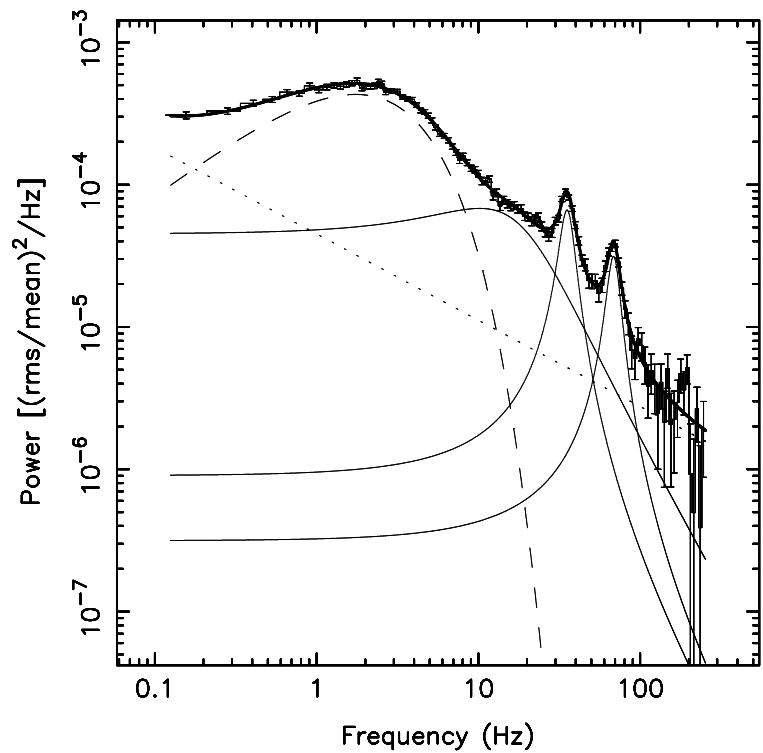}}
\caption{Example of a fit to the \sz=0.4--0.5 power spectrum, showing the relative contribution of the individual components. The combined fit is represented by the thick solid line, the three Lorentzians/QPOs by the thin solid lines, the cut-off power law/LFN by the dashed line, and the power law/VLFN by the dotted line. The Poisson level was subtracted. The reduced $\chi^2$ of this fit was 1.03 for 204 degrees of freedom.
\label{pds_comp_fig}} \end{figure}

\begin{figure}[t] \centerline{\includegraphics[width=15cm]{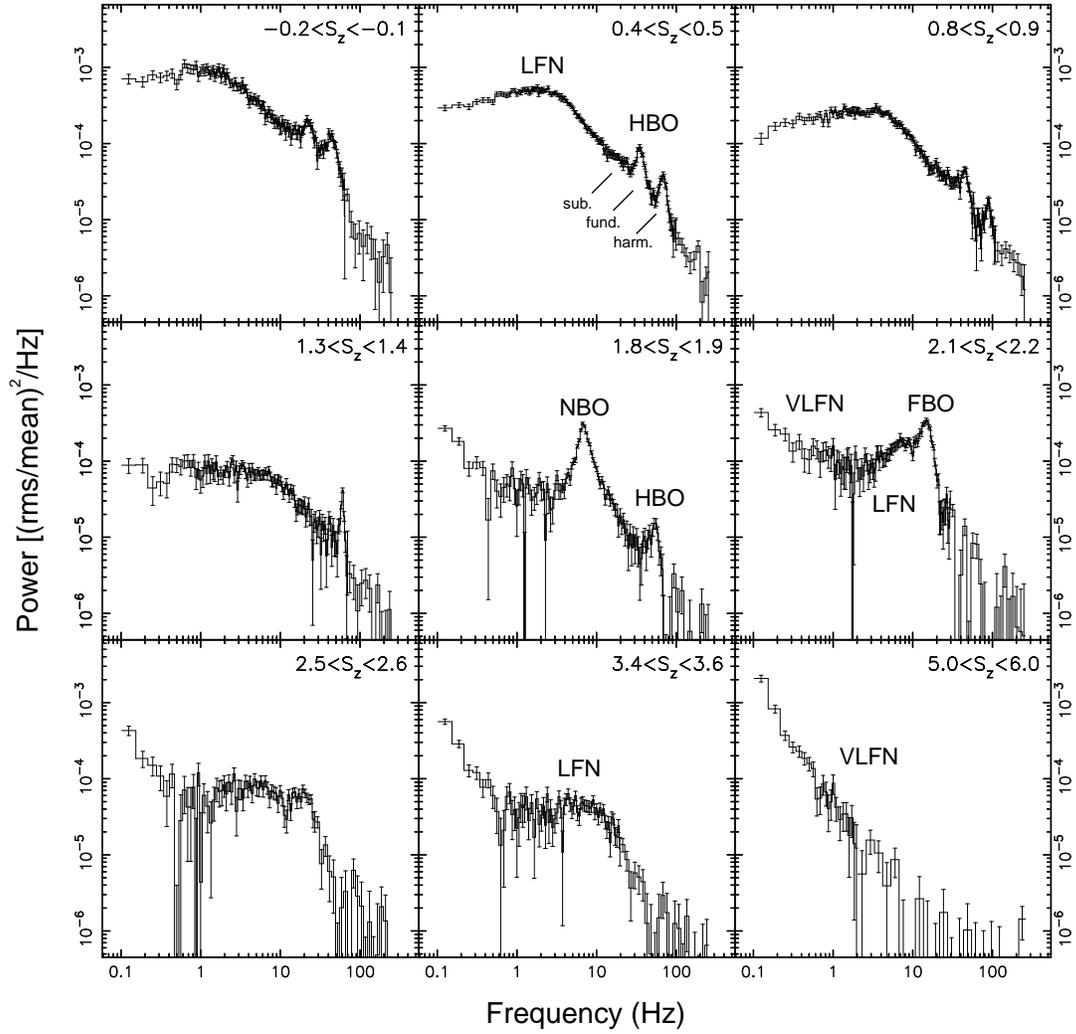}}
\caption{Power spectra (0.0625--256 Hz) for nine different \sz\,
selections. The Poisson level was subtracted for all power spectra.
The most important power spectral features are indicated.
\label{pds_fig}} \end{figure}

\begin{figure}[t] \centerline{\includegraphics[width=15cm]{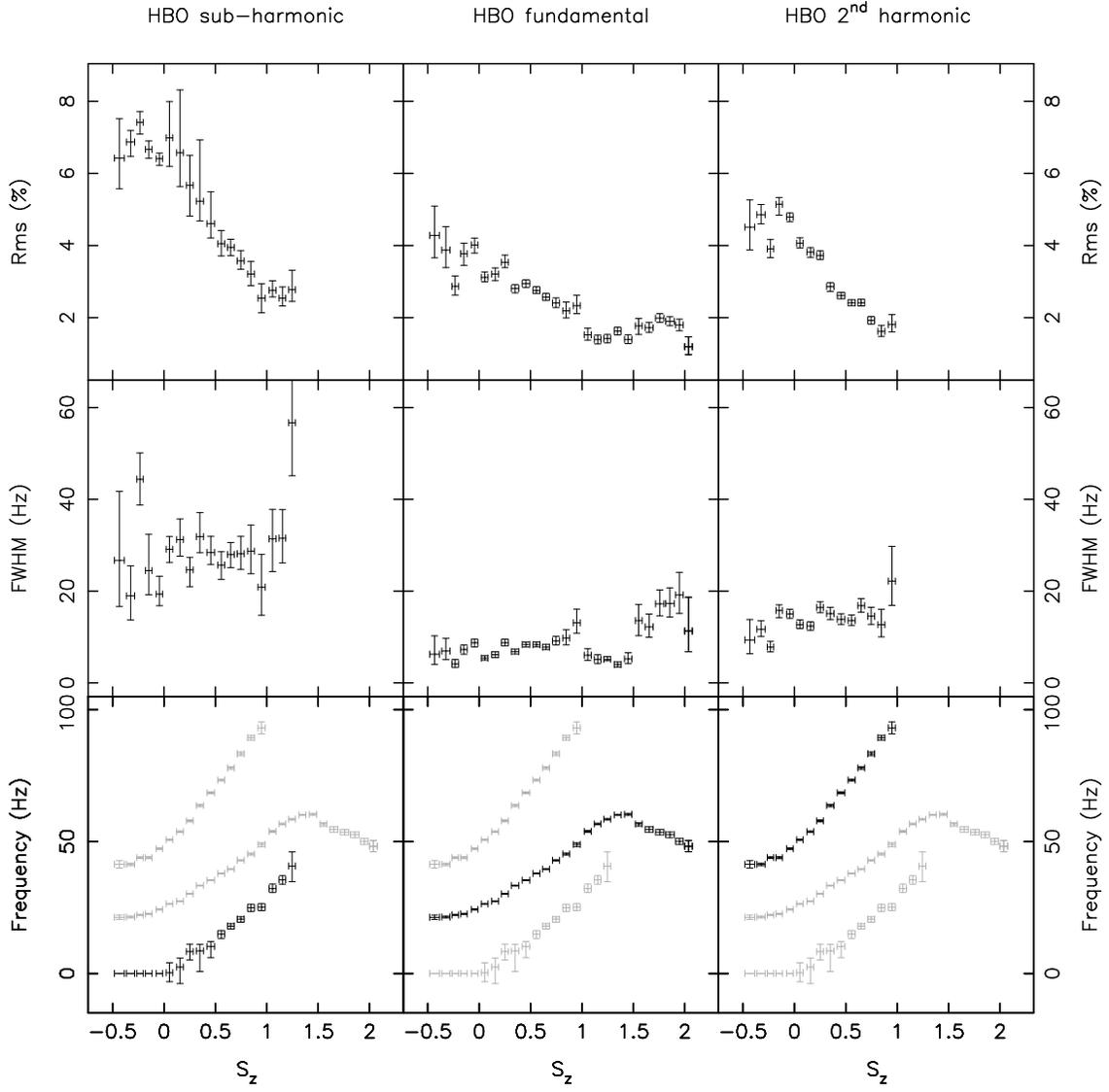}}
\caption{Properties of the HBO (middle column), its second
harmonic (right column), and sub-harmonic (left column) as a function
of \sz. For reasons of comparison the frequency of  the other two
QPOs (gray) are also plotted in the frequency plot of each QPO.
\label{hbo_fig}} \end{figure}

\begin{figure}[t] \centerline{\includegraphics[width=10cm]{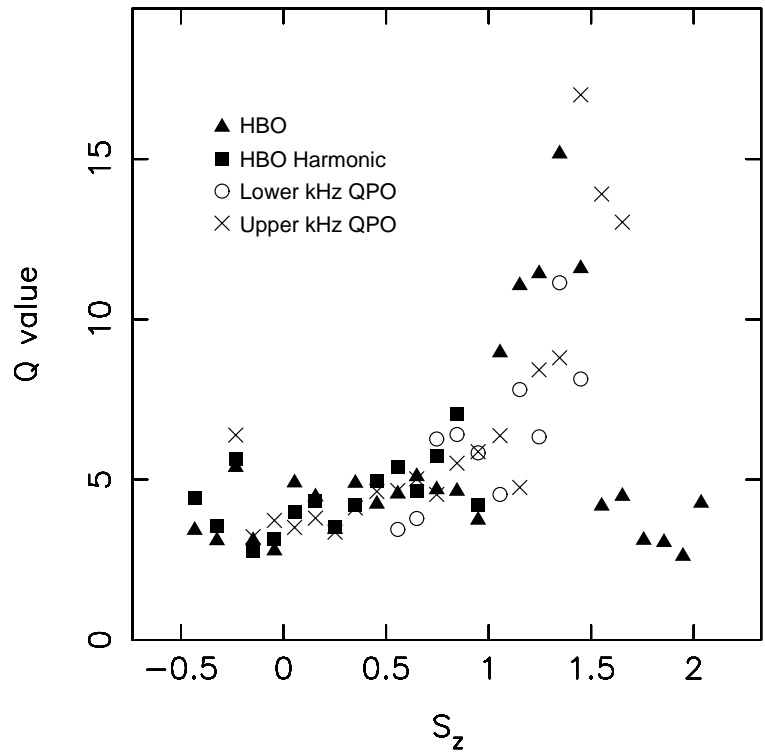}}
\caption{Q values (Frequency/FWHM) of the HBO, its
harmonic, and the two kHz QPOs as a function of \sz. For reasons of
clarity the error bars, which at a given \sz\ in general overlapped,
were omitted.\label{q-values_fig}} \end{figure}

\begin{figure}[t]
\centerline{\includegraphics[width=5cm]{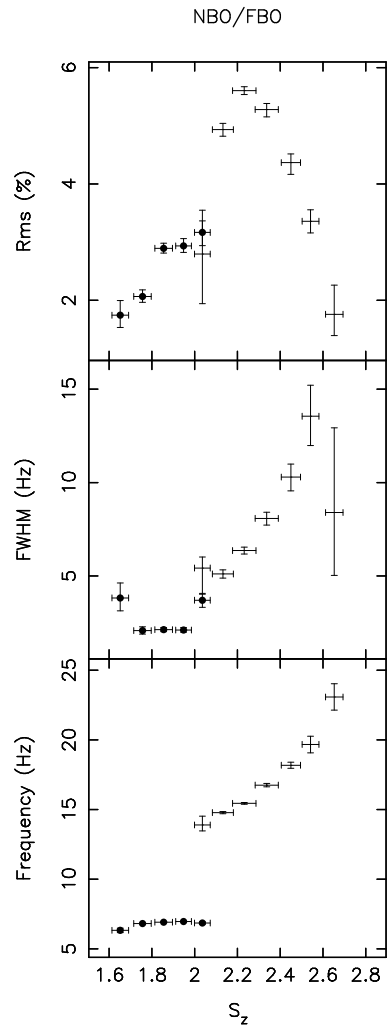}}
\caption{Properties of the NBO (filled circles) and FBO as a
function \sz.}\label{nbo_fig}
\end{figure}

\begin{figure}[t]
\centerline{\includegraphics[width=15cm,angle=-90]{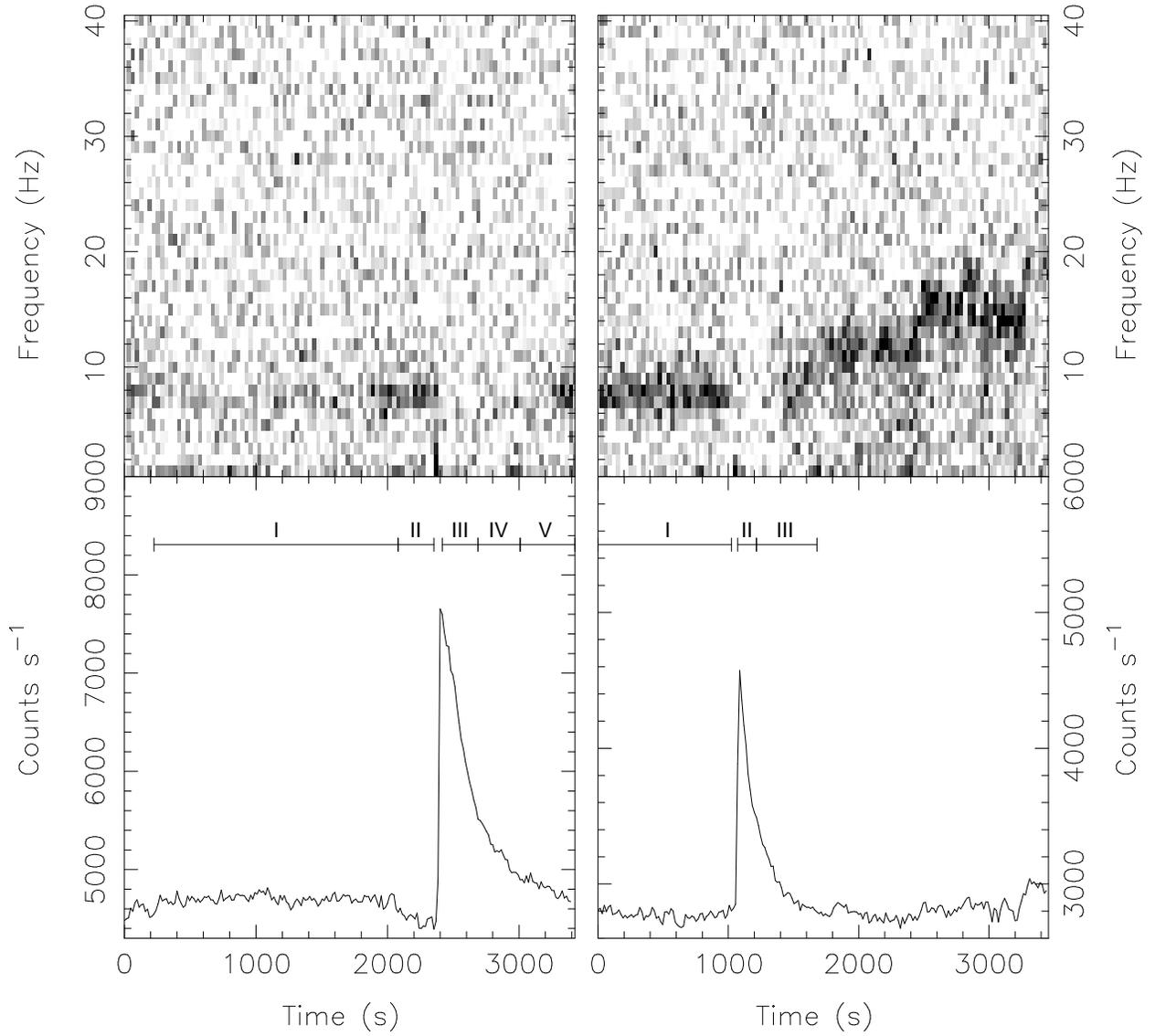}}
\caption{The dynamical power spectra {\it (top panels)} and light
curves in the 2--60 keV band {\it (bottom panels)} of the two type I
X-ray bursts in which we studied the behavior of the NBO/FBO. The
November 1998 burst is shown on the left (5 PCUs) and the October
1999, burst on the right (3 PCUs). The intervals in which we
measured the NBO/FBO are indicated by Roman numerals (see Table
\ref{burst-nbo_tab}). The shades of gray in the dynamical power
spectra represent the Leahy power, with darker shades indicating
higher powers.\label{burst-nbo_fig}} \end{figure}

\begin{figure}[t] \centerline{\includegraphics[width=10cm]{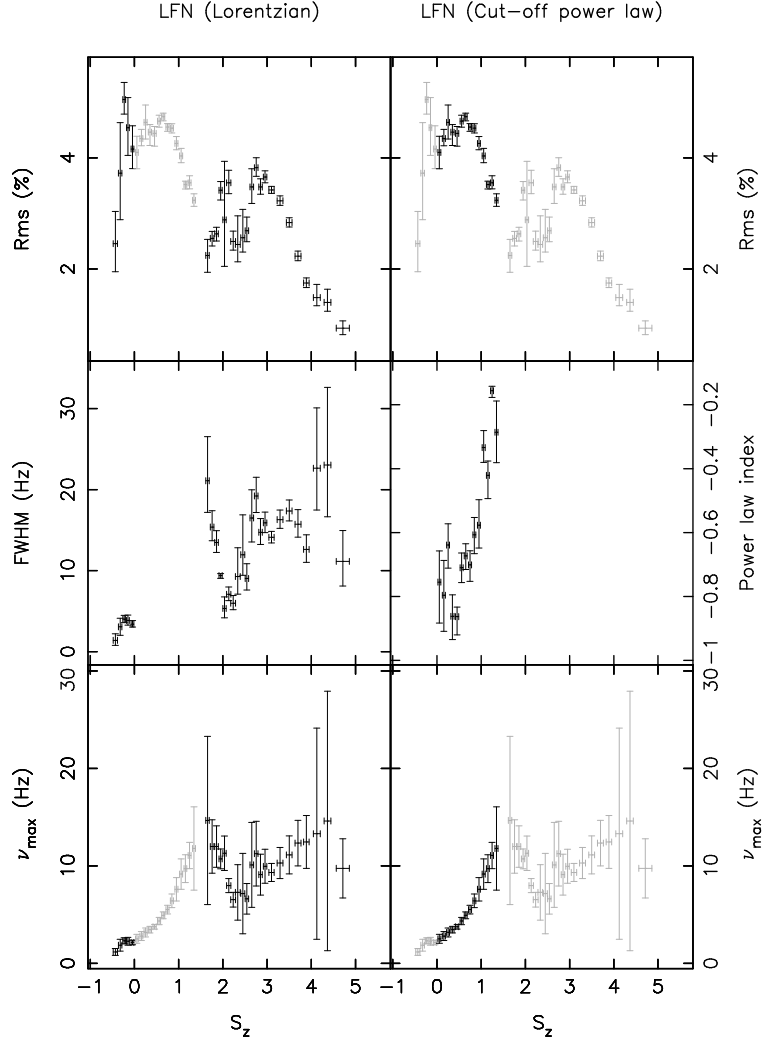}}
\caption{LFN properties as function of \sz. The left column
shows the results for the fits with a Lorentzian, the right column
those for fits with a cut-off power law. In both cases the rms
amplitude is the integrated power in the 1--100 Hz range. $\nu_{max}$
is the frequency at which most of the power is concentrated (see text
for the expressions for $\nu_{max}$.) For reasons of comparison we
also plotted the values of the other fit function (gray) in the
panels for the rms amplitude and $\nu_{max}$.\label{lfn_fig}}
\end{figure}

\begin{figure}[t] \centerline{\includegraphics[width=5cm]{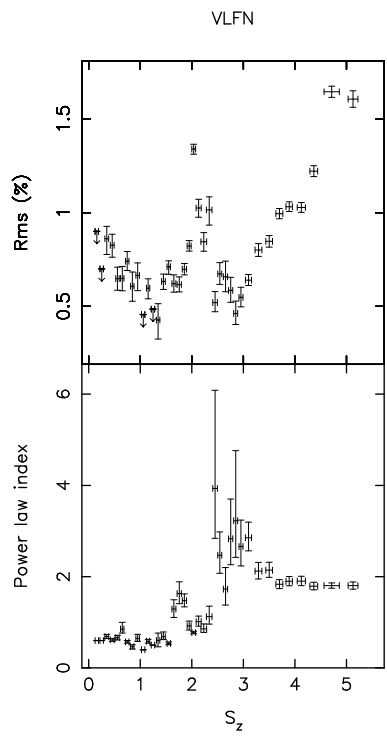}}
\caption{VLFN properties as function of \sz. The rms
amplitude is the integrated power in the 0.1--1 Hz range. The arrows
in the top panel represent upper limits.\label{vlfn_fig}} 
\end{figure}

\begin{figure}[t] 
\centerline{\includegraphics[width=10cm]{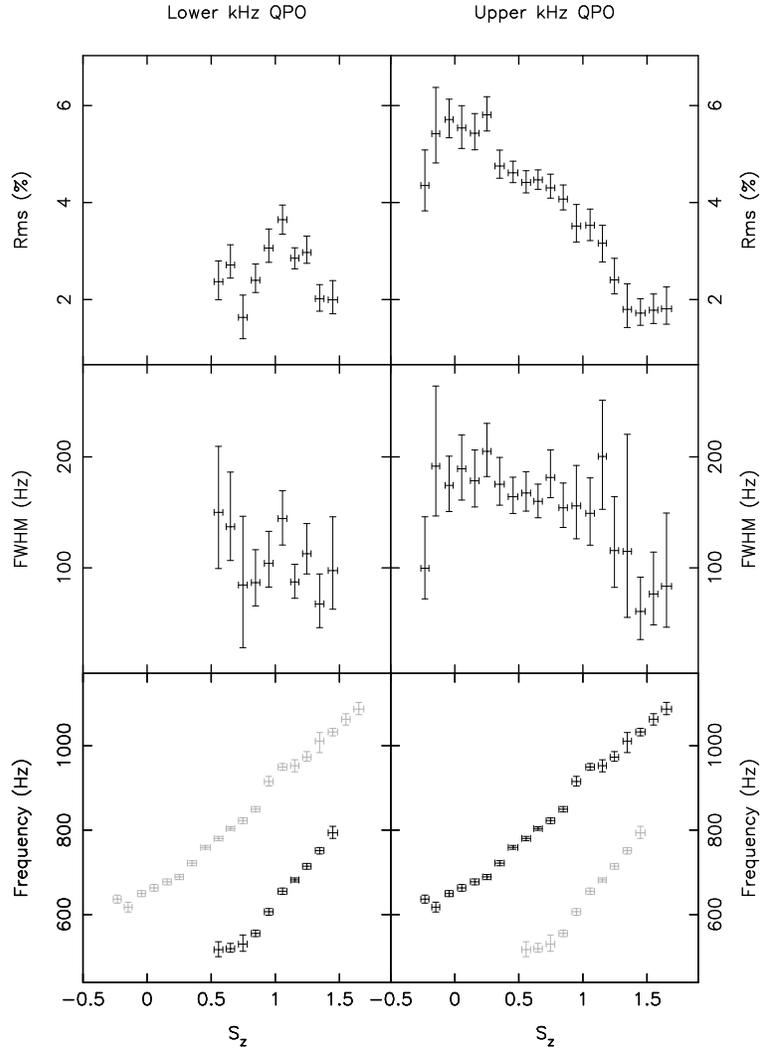}}
\caption{kHz QPO properties as a function of \sz. For
reasons of comparison, the  frequency of the other QPO (gray) is also
plotted in the frequency plot of each QPO.\label{khz_fig}}
\end{figure}

In our observations GX 17+2 traced out the well known Z-like tracks
in the CD and HID. In addition to motion along this track, we also
observed secular motion of the Z track itself in the HID of epoch 3
(see Figure \ref{cd-hid_fig}), as was already reported by
\citet{wihova1997} \citep[see also][for a possible shift in the
EXOSAT HID of GX 17+2]{kuvaoo1997}. Below \sz=1 the fluxes in the hard and soft energy bands behaved quite differently (Fig.\
\ref{sz-count_fig}); unlike the soft flux, the hard flux
continued the trend that is observed above \sz=1. Hence, the HB/NB
vertex in the CD is entirely due to the turn over of the soft
spectral flux. It is not clear what causes the soft spectral
flux to decrease below \sz=1. Although it is partly an
instrumental effect -- the peak of the soft component slowly shifts
out of the PCA effective energy range as it becomes softer-- most
likely intrinsic source changes occur at \sz=1. Most variability
components increase in strength below \sz=1, suggesting that they are
related to the hard spectral component. 

It is commonly believed that the motion of the source along and its
position on the Z track are determined by the mass accretion rate
($\dot{M}$) through the inner accretion disk and onto the neutron
star, increasing from the HB to the FB. Even though the count rates 
actually decrease on the NB and in some sources also on the FB
(implying an anti-correlation between $\dot{M}$ and flux) there are
several observational results that support this view: (1) the
frequencies of the HBO and kHz QPOs, that in many models are directly
related to $\dot{M}$, gradually increase from HB to NB \citep[and in
Sco X-1 even to the FB, see][]{vaswzh1996}. (2) The optical and UV
flux increase from HB to FB
\citep{havaeb1990,vrraga1990,vaalca1990,vrpera1991,hevawo1992,aukapa1992};
in X-ray binaries they are thought to be due to reprocessing  of
X-rays from the central source in the outer accretion disk.  There
are several reasons to believe that the  optical and UV flux are
betters tracers of the mass accretion rate than the X-ray flux
\citep{havaeb1990}. This would then solve the apparent
anti-correlation between $\dot{M}$ and the (X-ray) flux
\citep{havaeb1990}. (3) Motion of a source is always along the Z
track; jumps between branches are never observed, in accordance with
the assumption that $\dot{M}$ varies continuously. 

It has been found that secular motion of the Z tracks in the CDs of
the Cyg-like sources does not affect the relation between the rapid
variability properties and source position along the track (see, e.g.
\citealt{kuvaoo1994,jovawi2000}, however see \citealt{wiva2001}). In
view of the \mdot-driven picture mentioned above, these changes
can therefore not be due to changes in \mdot, but should have a
different origin. It has been suggested that the Cyg-like sources are
viewed at  a higher inclination than the Sco-like sources
\citep{kuvaoo1994}. The differences in shapes of the Z tracks and
some of the variability properties are also explained in this model
\citep{kuvaoo1994,kuva1995}. \citet{pslami1995} suggest that the
difference among the Z sources might be explained by a difference in
the magnetic field strength, with the Sco-like sources having a lower
magnetic field.

In the last few years results have been obtained that may challenge
the above described \mdot-driven picture. These results are mainly
due to the arrival of \xte\ and {\it BeppoSAX}, whose high quality
data revealed new phenomena and  allowed for a better comparison
between different classes of X-ray binaries. In the following we
discuss how some of our results of GX 17+2 conflict with the 
\mdot-driven picture.

The frequency of the HBO in GX17+2 starts to decrease when the source
is half way through the NB. Although there are several explanations
for this (see Section \ref{disc-hbo_sec}), it means that the first
assumption given above (i.e.\ QPO frequencies increase with \mdot) is
not as solid as it seemed a few years ago. 

The theory of type I X-ray bursts predicts that their properties
change and correlate with \mdot\  \citep{fuhami1981,fula1987}. A
study of the type I X-ray bursts in GX 17+2 by \citet{kuhova2001}
showed that their properties did not correlate well with the position
along the Z track at which they occurred, and therefore not well with
the inferred \mdot. Although other explanations are possible
\citep[see][]{kuhova2001}, this could mean that \mdot\ does not
determine the position of the source along the Z track. However, note
that there does seem to be a correlation between burst occurrence and
position along the Z track, which is usually explained by a changes
in \mdot\ onto the neutron star (along the Z track).

Recently, \citet{distro2000} reported the discovery with BeppoSAX of
a hard  tail in the energy spectrum of GX 17+2. It increased in
strength from the HB/NB vertex (where only upper limits on its
strength could be determined) onto the HB. They also showed that the
total 0.1--200 keV flux increased monotonically from the NB/FB vertex
($F_X=1.52\times10^{-8}\,\,erg\,s^{-1}$) to the top of the HB
($F_X=1.84\times10^{-8}\,\,erg\,s^{-1}$). This is different from the
behavior of the 2.9--20.1 keV count rate (see Figure
\ref{sz-count_fig}a), which increased from NB/FB vertex to the HB/NB
vertex, but decreases afterwards.  As can be seen from Figure
\ref{sz-count_fig}b the behavior of the hard count rate confirms the
findings of \citet{distro2000}. Also, there is no evidence for a
change in its behavior near the HB/NB vertex. Unfortunately no
BeppoSAX data were taken on the FB to compare to our RXTE data. Hard
power law tails have also been found in the Z sources GX 5-1
\citep{asdomi1994}, Cyg X-2 \citep{frdama1998,difabu2001}, Sco X-1
\citep{stba2000,dahero2001} and GX 349+2 \citep{diroia2001}. Except
for Sco X-1, where no clear correlation with spectral state was found
\citep{dahero2001}, in all these sources the strength of the hard
power law tail seems to decrease in the same direction as was
observed in GX 17+2, i.e. from HB to FB.  

The behavior of the 0.1--200 keV flux on the NB and HB suggest that
\mdot\ actually increases in the opposite direction from what is
commonly assumed. However, since the flux changes are rather moderate
($\sim$20\%), other options are that \mdot\ does not change at all
along the Z track, or at least that it shows no good correlation with
the position along the Z. In that case motion along the Z track (i.e.
spectral change) is caused by another parameter, for example the
inner disk radius \citep{va2000}. A similar possibility was suggested
by \citet{howiva2001} to explain the observed behavior of  the black
hole transient XTE J1550--564. They found that spectral transitions
occurred at different levels of the inferred \mdot. The spectral
changes were ascribed to an unknown parameter, which they suggested
to be the inner disk radius (based on variability), the size of the
Comptonizing region (based on spectral hardness), and/or the
accretion flow geometry (based on radio brightness). Note that
options like the latter one may lead to changes in the \mdot\ onto
the neutron star, while the \mdot\ through the inner disk remains
constant; in that way the occurrence of type X-rays bursts might
become dependent on the position along the Z track.  If the above
also applies to the Z sources, the secular motion that is observed
could perhaps be explained by changes in the \mdot. The apparent lack
of a correlation between the type I burst properties and the position
along the Z track could then also be explained, since \mdot\ would
determine the properties of the bursts but no longer the spectral
properties. 

It is interesting to see that GX 17+2 and the Z sources, show in fact
several similarities with XTE J1550--564 and other black hole
sources. The first was already briefly mentioned above, and concerns
the independent behavior of the hard and soft flux on the HB. This
independence of  the hard and soft spectral components is commonly
observed in black hole candidates \citep[BHCs, see][]{tale1995}.
Moreover, there are several properties that correlate well with the
spectral hardness in BHCs and also with the spectral hardness in Z
sources, which increases from the NB to the HB (for the moment we
will exclude the FB from the comparison; it will be discussed later).
These properties are the radio flux, the optical flux, and to a
certain degree also QPO frequencies. (1) The radio flux in BHCs is 
correlated with the spectral hardness \citep[and references
therein]{fecotz1999,fe2001}. This is also true for GX 17+2, where the
radio flux increases from the NB to the HB \citep{pelezi1988}. Sco
X-1 \citep{hjstwh1990} and Cyg X-2 \citep{hjhaco1990} show similar
behavior, whereas GX 5-1 only showed one radio flare on the NB
\citep{talehj1992}. The behavior of the radio flux suggests that as
the spectral hardness increases, part of the flow onto the neutron
star evolves into an outflow. (2) \cite{jabaor2001} found that in XTE
J1550--564 the optical flux decreased during several small hard X-ray
flares. Although the optical counterpart of GX 17+2
\citep{cafiga1999,demaan1999} has never been studied in much detail,
this behavior is similar to the anti-correlation between the hard
color  and optical/UV flux on the NB and HB in the Z sources  Cyg X-2
\citep{havaeb1990,vrraga1990,vaalca1990} and Sco X-1
\citep{vrpera1991,hevawo1992,aukapa1992}.  (3) In XTE J1550--564 the
frequencies of the low and high frequency QPO anti-correlate with the
hard flux \citep{somcre2000,howiva2001}. This is also found in GX
17+2 for the kHz QPOs and HBO (compare Figure \ref{sz-count_fig}b
with Figures \ref{hbo_fig} and \ref{khz_fig}) and other Z sources.
The only exception to this, together with the odd behavior of the
hard tail in Sco X-1 (see above) is the correlation between the HBO
frequency and spectral hardness that we observed on the lower NB. In
that respect it is interesting to mention that  \citet{ruleva1999}
report a switch from a correlation to an anti-correlation between the
QPO frequency and spectral hardness in several BHCs, when the latter
increases beyond a certain value. In addition to the QPO frequencies
also the band-limited noise behaves similarly. It becomes stronger
with spectral hardness, both in BHCs and Z sources \citep{va1995a}

It is tempting to compare the different black hole states
\citep{tale1995,va1995a} with those of the Z sources. Based on the
above similarities, we suggest that the HB corresponds to the low
state, the NB to the intermediate and/or very high state state, and
the FB to the high state. However, there are still considerable
differences: the low state spectra are much harder than those of the
HB, and also the noise in the low state is much stronger. Moreover,  
flaring is not observed in the black hole high state. However, in
both the high state and FB the variability at high frequencies ($>$50
Hz) is extremely weak. It is not clear if and how the presence of a
solid surface can account for these differences.

\section{Summary and conclusions}

Our  most important findings are the variable frequency difference of
the kHz QPO, the non-monotonic relation between the HBO and the upper
kHz QPO, the similar Q values of the HBO, its second harmonic, and
the kHz QPOs and the suppression of the NBO amplitude during type I
X-ray bursts. 

In the following we summarize what the implications of each of our
new findings are for some of the proposed models. The models we
considered for the HBO and kHz QPOs were the relativistic precession
model (RPM), the 'disk-filter' model, the sonic point beat frequency
mode (SPBFM), and the two-oscillator model. The latter model can be
discarded on the basis of our results \citep[see][]{jovaho2001}. For
the disk-filter model we assumed that it produces the same
frequencies as the RPM and it will therefore not be discussed it
separately.

(1) HBO-kHz QPO frequencies: The RPM can only explain a turnover of
the HBO-kHz QPO frequency relation when one assumes that the observed
HBO is four times the predicted frequency and then only for  rather
extreme neutron star parameters: a high neutron star spin frequency
($\sim$500--700 Hz), a neutron star mass of more than 2 $M_\odot$,
and a hard equation of state. The SPBFM needs some extension; for
example, decoupling of part of the accretion flow from the disk flow
outside the magnetosphere (to explain the HBO frequency decrease) and
a subsequent recoupling of this flow with the disk flow before the
sonic point radius is reached (to explain the increase in the kHz QPO
frequencies). How this should exactly work is not clear.

\begin{figure}[t]
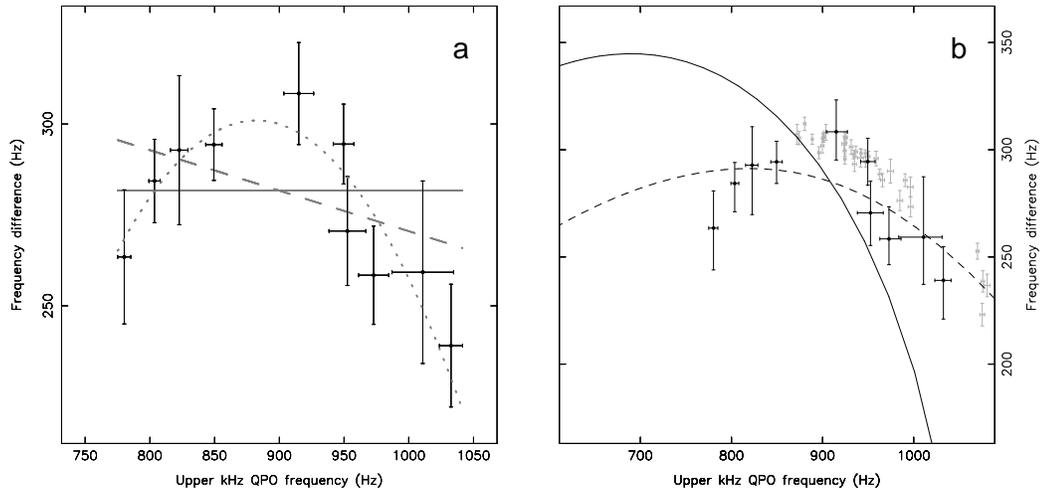
 
\centerline{
\includegraphics[width=6.5cm]{f13a.ps}
\includegraphics[width=6.5cm]{f13b.ps}} 

\caption{The kHz QPO frequency difference in GX 17+2 as function of
the upper peak frequency. In (a)  three fits to the data are shown
(gray): a constant (solid line), a straight line (dashed line), and a
parabola (dotted line). In (b) the two best fits to the data are
shown, for theoretical relations between the radial epicyclic
frequency and azimuthal frequency for circular orbtis (solid line)
and eccentric orbits (dashed line). The fits were kindly provided by
Draza Markovi\'c.  Note that the dashed line requires highly
eccentric orbits with apastron to periastron ratios of $\sim$2--3.
For comparison we also plotted the data for Sco X-1 (gray) from
\citet{vawiho1997}. For additional details see
text\label{khz-diff_fig}} 

\end{figure}

\begin{figure}[t]
\centerline{\includegraphics[width=10cm]{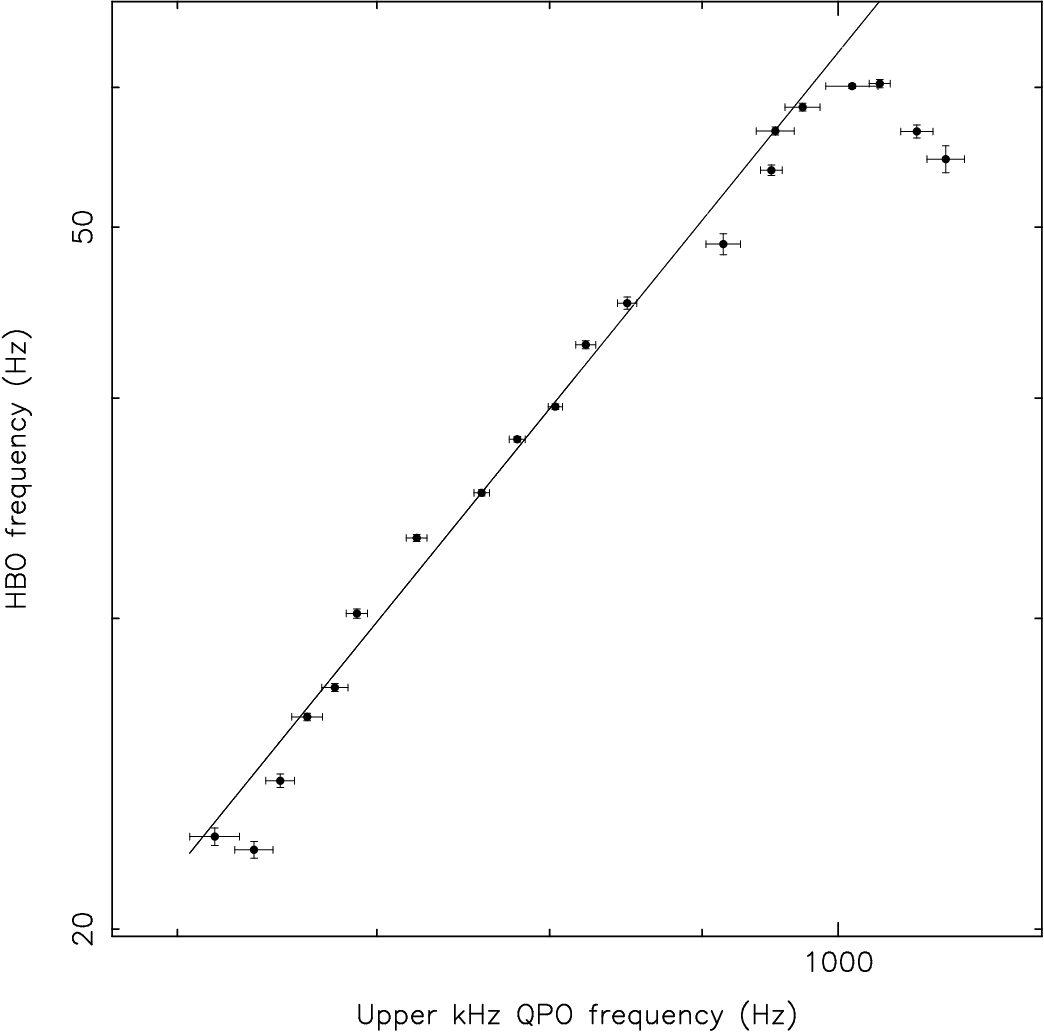}}
\caption{HBO frequency as function upper kHz QPO frequency.
The solid line is the best power law fit to the data below 1000 Hz
for the upper kHz frequency. The power law index is 2.08$\pm$0.07.
A clear deviation from this relation can be seen for values above 1000
Hz.\label{khz-hbo_fig}} \end{figure}

(2) Variable kHz QPO separation: Although not significant at a 3
sigma level, the kHz QPO separation seems to decrease both towards
low and high values for the upper kHz QPO frequency. The RPM
basically predicts such behavior, but in its current form it is
inconsistent with our results, as is apparent from  Figure
\ref{khz-diff_fig}. The SPBFM can not explain the possible decrease
of the frequency separation at the low frequency side.

(3) Similar Q values: In the SPBFM  the  low and high frequency
QPOs are produced independently from each other and there is no a
priori reason why their Q values should be similar. In the RPM, all
the oscillations are produced by a single test particle. Lifetime
broadening and the presence of multiple test particles both lead to
different Q values for the low and high frequency QPO. Frequency
modulation could work for both models, but only if the modulating
parameter is not the same as the one that causes the motion along the
Z track.

It is clear that both the RPM and SPBFM in their current forms cannot 
satisfactorily explain all the observed phenomena. 

The observed behavior of the NBO during two type I X-ray bursts
showed that this QPO is rather sensitive to the strength of the
radiation field from the neutron star. This seems to confirm the
basic idea behind the feedback mechanism model that was proposed by
\citet{folami1989}, i.e. it requires a delicate balance between the
inwards directed forces and the radiation field.

Finally, we compared the behavior of GX 17+2 (and the other Z
sources) with that of (some) black hole LMXBs. Many similarities
exist, particularly with regard to the relation between the hard
spectral component on the one hand, and variability properties (QPO
frequency, broad band noise strength) and radio flux on the other.
Based on recent findings in black hole LMXBs we conclude that the
mass accretion might not be the parameter that determines the
position along the Z track after all.

\acknowledgments \noindent This work was supported in part by the
Netherlands Organization for Scientific Research (NWO) grant
614-51-002 and by NWO Spinoza grant 08-0 to E.P.J. van den Heuvel. 
RW was supported by NASA through the Chandra Postdoctoral Fellowship
grant number PF9-10010 awarded by the Chandra X-ray Center, which is
operated by the Smithsonian Astrophysical Observatory for NASA under
contract NAS8-39073. MM is a fellow of the Consejo Nacional de
Investigaciones Cient\'{\i}ficas y T\'ecnicas de la Rep\'ublica
Argentina. WHGL is grateful for support from NASA. The authors thank
the referee for his/her thorough reading and usefull remarks.
Finally, we would like to thank Draza Markovi\'c for helpfull
comments and for providing theoretical fits to our data.


\begin{thebibliography}{92}
\expandafter\ifx\csname natexlab\endcsname\relax\def\natexlab#1{#1}\fi

\bibitem[{{Alpar} {et~al.}(1992){Alpar}, {Hasinger}, {Shaham}, \&
  {Yancopoulos}}]{alhash1992}
{Alpar}, M.~A., {Hasinger}, G., {Shaham}, J., \& {Yancopoulos}, S. 1992, \aap,
  257, 627

\bibitem[{{Alpar} \& {Shaham}(1985)}]{alsh1985}
{Alpar}, M.~A. \& {Shaham}, J. 1985, \nat, 316, 239

\bibitem[{{Asai} {et~al.}(1994){Asai}, {Dotani}, {Mitsuda}, {Nagase}, {Kamado},
  {Kuulkers}, \& {Breedon}}]{asdomi1994}
{Asai}, K., {Dotani}, T., {Mitsuda}, K., {et~al.} 1994, \pasj, 46, 479

\bibitem[{{Augusteijn} {et~al.}(1992){Augusteijn}, {Karatasos}, {Papadakis},
  {Paterakis}, {Kikuchi}, {Brosch}, {Leibowitz}, {Hertz}, {Mitsuda}, {Dotani},
  {Lewin}, {van del Klis}, \& {van Paradijs}}]{aukapa1992}
{Augusteijn}, T., {Karatasos}, K., {Papadakis}, M., {et~al.} 1992, \aap, 265,
  177

\bibitem[{{Belloni} {et~al.}(1997){Belloni}, {van der Klis}, {Lewin}, {van
  Paradijs}, {Dotani}, {Mitsuda}, \& {Miyamoto}}]{bevale1997}
{Belloni}, T., {van der Klis}, M., {Lewin}, W. H.~G., {et~al.} 1997, \aap, 322,
  857

\bibitem[{{Berger} \& {van der Klis}(1994)}]{beva1994}
{Berger}, M. \& {van der Klis}, M. 1994, \aap, 292, 175

\bibitem[{{Bradt} {et~al.}(1993){Bradt}, {Rothschild}, \& {Swank}}]{brrosw1993}
{Bradt}, H.~V., {Rothschild}, R.~E., \& {Swank}, J.~H. 1993, \aaps, 97, 355

\bibitem[{{Callanan} {et~al.}(1999){Callanan}, {Filippenko}, \&
  {Garcia}}]{cafiga1999}
{Callanan}, P.~J., {Filippenko}, A.~V., \& {Garcia}, M.~R. 1999, \iaucirc, 7219

\bibitem[{{D'Amico} {et~al.}(2001){D'Amico}, {Heindl}, {Rothschild}, \&
  {Gruber}}]{dahero2001}
{D'Amico}, F., {Heindl}, W.~A., {Rothschild}, R.~E., \& {Gruber}, D.~E. 2001,
  \apjl, 547, L147

\bibitem[{{Deutsch} {et~al.}(1999){Deutsch}, {Margon}, {Anderson}, {Wachter},
  \& {Goss}}]{demaan1999}
{Deutsch}, E.~W., {Margon}, B., {Anderson}, S.~F., {Wachter}, S., \& {Goss},
  W.~M. 1999, \apj, 524, 406

\bibitem[{{di Salvo} {et~al.}(2001{\natexlab{a}}){di Salvo}, {Farinelli},
  {Burderi}, {Frontera}, {Kuulkers}, {Masetti}, {Robba}, {Stella}, \& {van der
  Klis}}]{difabu2001}
{di Salvo}, T., {Farinelli}, R., {Burderi}, L., {et~al.} 2001{\natexlab{a}},
  \aap, submitted

\bibitem[{{di Salvo} {et~al.}(2001{\natexlab{b}}){di Salvo}, {Robba}, {Iaria},
  {Stella}, {Burderi}, \& {Israel}}]{diroia2001}
{di Salvo}, T., {Robba}, N.~R., {Iaria}, R., {et~al.} 2001{\natexlab{b}}, \apj,
  554, 49

\bibitem[{{di Salvo} {et~al.}(2000){di Salvo}, {Stella}, {Robba}, {van der
  Klis}, {Burderi}, {Israel}, {Homan}, {Campana}, {Frontera}, \&
  {Parmar}}]{distro2000}
{di Salvo}, T., {Stella}, L., {Robba}, N.~R., {et~al.} 2000, \apjl, 544, L119

\bibitem[{{Dieters} \& {van der Klis}(2000)}]{diva2000}
{Dieters}, S.~W. \& {van der Klis}, M. 2000, \mnras, 311, 201

\bibitem[{{Fender}(2001)}]{fe2001}
{Fender}, R. 2001, To be published in Proc. ESO workshop `Black Holes in
  binaries and galactic nuclei', Eds L. Kaper, E.P.J. van den Heuvel and P.A.
  Woudt, Springer-Verlag, astro-ph/9911176

\bibitem[{{Fender} {et~al.}(1999){Fender}, {Corbel}, {Tzioumis}, {McIntyre},
  {Campbell-Wilson}, {Nowak}, {Sood}, {Hunstead}, {Harmon}, {Durouchoux}, \&
  {Heindl}}]{fecotz1999}
{Fender}, R., {Corbel}, S., {Tzioumis}, T., {et~al.} 1999, \apjl, 519, L165

\bibitem[{{Ford} {et~al.}(1998){Ford}, {van der Klis}, {van Paradijs},
  {M{\'e}ndez}, {Wij\-nands}, \& {Kaaret}}]{fovava1998}
{Ford}, E.~C., {van der Klis}, M., {van Paradijs}, J., {et~al.} 1998, \apjl,
  508, L155

\bibitem[{{Fortner} {et~al.}(1989){Fortner}, {Lamb}, \& {Miller}}]{folami1989}
{Fortner}, B., {Lamb}, F.~K., \& {Miller}, G.~S. 1989, \nat, 342, 775

\bibitem[{{Frontera} {et~al.}(1998){Frontera}, {dal Fiume}, {Malaguti},
  {Nicastro}, {Orlandini}, {Palazzi}, {Pian}, {Favata}, \&
  {Santangelo}}]{frdama1998}
{Frontera}, F., {dal Fiume}, D., {Malaguti}, G., {et~al.} 1998, in The Active
  X-ray Sky: Results from BeppoSAX and RXTE, p286

\bibitem[{{Fujimoto} {et~al.}(1981){Fujimoto}, {Hanawa}, \&
  {Miyaji}}]{fuhami1981}
{Fujimoto}, M.~Y., {Hanawa}, T., \& {Miyaji}, S. 1981, \apj, 267

\bibitem[{{Fushiki} \& {Lamb}(1987)}]{fula1987}
{Fushiki}, I. \& {Lamb}, D.~Q. 1987, \apjl, 323, L55

\bibitem[{{Hasinger} \& {van der Klis}(1989)}]{hava1989}
{Hasinger}, G. \& {van der Klis}, M. 1989, \aap, 225, 79

\bibitem[{{Hasinger} {et~al.}(1990){Hasinger}, {van der Klis}, {Ebisawa},
  {Dotani}, \& {Mitsuda}}]{havaeb1990}
{Hasinger}, G., {van der Klis}, M., {Ebisawa}, K., {Dotani}, T., \& {Mitsuda},
  K. 1990, \aap, 235, 131

\bibitem[{{Hertz} {et~al.}(1992){Hertz}, {Vaughan}, {Wood}, {Norris},
  {Mitsuda}, {Michelson}, \& {Dotani}}]{hevawo1992}
{Hertz}, P., {Vaughan}, B., {Wood}, K.~S., {et~al.} 1992, \apj, 396, 201

\bibitem[{{Hjellming} {et~al.}(1990{\natexlab{a}}){Hjellming}, {Han},
  {Cordova}, \& {Hasinger}}]{hjhaco1990}
{Hjellming}, R.~M., {Han}, X.~H., {Cordova}, F.~A., \& {Hasinger}, G.
  1990{\natexlab{a}}, \aap, 235, 147

\bibitem[{{Hjellming} {et~al.}(1990{\natexlab{b}}){Hjellming}, {Stewart},
  {White}, {Strom}, {Lewin}, {Hertz}, {Wood}, {Norris}, {Mitsuda}, {Penninx},
  \& {van Paradijs}}]{hjstwh1990}
{Hjellming}, R.~M., {Stewart}, R.~T., {White}, G.~L., {et~al.}
  1990{\natexlab{b}}, \apj, 365, 681

\bibitem[{{Homan}(2001)}]{ho2001}
{Homan}, J. 2001, in X-ray timing studies of low-mass X-ray binaries, PhD
  Thesis, Chapter 9

\bibitem[{{Homan} {et~al.}(1999){Homan}, {Jonker}, {Wij\-nands}, {van der
  Klis}, \& {van Paradijs}}]{hojowi1999}
{Homan}, J., {Jonker}, P.~G., {Wij\-nands}, R., {van der Klis}, M., \& {van
  Paradijs}, J. 1999, \apjl, 516, L91

\bibitem[{{Homan} {et~al.}(2001){Homan}, {Wij\-nands}, {van der Klis},
  {Belloni}, {van Paradijs}, {Klein-Wolt}, {Fender}, \&
  {M\'endez}}]{howiva2001}
{Homan}, J., {Wij\-nands}, R., {van der Klis}, M., {et~al.} 2001, \apjs, 132,
  in press

\bibitem[{{Jahoda} {et~al.}(1996){Jahoda}, {Swank}, {Giles}, {Stark},
  {Strohmayer}, {Zhang}, \& {Morgan}}]{jaswgi1996}
{Jahoda}, K., {Swank}, J.~H., {Giles}, A.~B., {et~al.} 1996, \procspie, 2808,
  59

\bibitem[{{Jain} {et~al.}(2001){Jain}, {Bailyn}, {Orosz}, {McClintock},
  {Sobczak}, \& {Remillard}}]{jabaor2001}
{Jain}, R.~J., {Bailyn}, C.~D., {Orosz}, J.~A., {et~al.} 2001, \apj, in press

\bibitem[{{Jonker} {et~al.}(2001){Jonker}, {van der Klis}, {Homan},
  {M{\'e}ndez}, {Wij\-nands}, {Lewin}, \& {Zhang}}]{jovaho2001}
{Jonker}, P.~G., {van der Klis}, M., {Homan}, J., {et~al.} 2001, \mnras,
  submitted

\bibitem[{{Jonker} {et~al.}(2000{\natexlab{a}}){Jonker}, {van der Klis},
  {Homan}, {Wij\-nands}, {van Paradijs}, {M{\'e}ndez}, {Kuul\-kers}, \&
  {Ford}}]{jovaho2000}
---. 2000{\natexlab{a}}, \apj, 531, 453

\bibitem[{{Jonker} {et~al.}(1999){Jonker}, {van der Klis}, \&
  {Wij\-nands}}]{jovawi1999}
{Jonker}, P.~G., {van der Klis}, M., \& {Wij\-nands}, R. 1999, \apjl, 511, L41

\bibitem[{{Jonker} {et~al.}(2000{\natexlab{b}}){Jonker}, {van der Klis},
  {Wij\-nands}, {Homan}, {van Paradijs}, {M{\'e}ndez}, {Ford}, {Kuul\-kers}, \&
  {Lamb}}]{jovawi2000}
{Jonker}, P.~G., {van der Klis}, M., {Wij\-nands}, R., {et~al.}
  2000{\natexlab{b}}, \apj, 537, 374

\bibitem[{{Jonker} {et~al.}(1998){Jonker}, {Wij\-nands}, {van der Klis},
  {Psaltis}, {Kuul\-kers}, \& {Lamb}}]{jowiva1998}
{Jonker}, P.~G., {Wij\-nands}, R., {van der Klis}, M., {et~al.} 1998, \apjl,
  499, L191

\bibitem[{{Kuul\-kers} {et~al.}(2001){Kuul\-kers}, {Homan}, {van der Klis},
  {Lewin}, \& {M\'endez}}]{kuhova2001}
{Kuul\-kers}, E., {Homan}, J., {van der Klis}, M., {Lewin}, W.~H.~G., \&
  {M\'endez}, M. 2001, \aap, submitted

\bibitem[{{Kuul\-kers} \& {van der Klis}(1995)}]{kuva1995}
{Kuul\-kers}, E. \& {van der Klis}, M. 1995, \aap, 303, 801

\bibitem[{{Kuul\-kers} \& {van der Klis}(1996)}]{kuva1996}
---. 1996, \aap, 314, 567

\bibitem[{{Kuul\-kers} {et~al.}(1994){Kuul\-kers}, {van der Klis},
  {Oosterbroek}, {Asai}, {Dotani}, {van Paradijs}, \& {Lewin}}]{kuvaoo1994}
{Kuul\-kers}, E., {van der Klis}, M., {Oosterbroek}, T., {et~al.} 1994, \aap,
  289, 795

\bibitem[{{Kuul\-kers} {et~al.}(1997){Kuul\-kers}, {van der Klis},
  {Oosterbroek}, {van Paradijs}, \& {Lewin}}]{kuvaoo1997}
{Kuul\-kers}, E., {van der Klis}, M., {Oosterbroek}, T., {van Paradijs}, J., \&
  {Lewin}, W. H.~G. 1997, \mnras, 287, 495

\bibitem[{{Kuul\-kers} {et~al.}(1996){Kuul\-kers}, {van der Klis}, \&
  {Vaughan}}]{kuvava1996}
{Kuul\-kers}, E., {van der Klis}, M., \& {Vaughan}, B.~A. 1996, \aap, 311, 197

\bibitem[{{Lamb} \& {Miller}(2001)}]{lami2001}
{Lamb}, F.~K. \& {Miller}, M.~C. 2001, \apjl, in preparation, astro-ph/0007460

\bibitem[{{Lamb} {et~al.}(1985){Lamb}, {Shibazaki}, {Alpar}, \&
  {Shaham}}]{lashal1985}
{Lamb}, F.~K., {Shibazaki}, N., {Alpar}, M.~A., \& {Shaham}, J. 1985, \nat,
  317, 681

\bibitem[{{Lewin} {et~al.}(1992){Lewin}, {Lubin}, {Tan}, {van der Klis}, {van
  Paradijs}, {Penninx}, {Dotani}, \& {Mitsuda}}]{leluta1992}
{Lewin}, W. H.~G., {Lubin}, L.~M., {Tan}, J., {et~al.} 1992, \mnras, 256, 545

\bibitem[{{Markovic} \& {Lamb}(2001)}]{mala2001}
{Markovic}, D. \& {Lamb}, F.~K. 2001, \mnras, submitted, astro-ph/0009169

\bibitem[{{M{\'e}ndez} \& {van der Klis}(1999)}]{meva1999}
{M{\'e}ndez}, M. \& {van der Klis}, M. 1999, \apjl, 517, L51

\bibitem[{{M\'endez} {et~al.}(1998{\natexlab{a}}){M\'endez}, {van der Klis},
  {van Paradijs}, {Lewin}, {Vaughan}, {Kuul\-kers}, {Zhang}, {Lamb}, \&
  {Psaltis}}]{mevava1998}
{M\'endez}, M., {van der Klis}, M., {van Paradijs}, J., {et~al.}
  1998{\natexlab{a}}, \apjl, 494, L65

\bibitem[{{M\'endez} {et~al.}(1998{\natexlab{b}}){M\'endez}, {van der Klis},
  {Wij\-nands}, {Ford}, {van Paradijis}, \& {Vaughan}}]{mevawi1998}
{M\'endez}, M., {van der Klis}, M., {Wij\-nands}, R., {et~al.}
  1998{\natexlab{b}}, \apjl, 505, L23

\bibitem[{{Miller} {et~al.}(1998){Miller}, {Lamb}, \& {Psaltis}}]{milaps1998}
{Miller}, M.~C., {Lamb}, F.~K., \& {Psaltis}, D. 1998, \apj, 508, 791

\bibitem[{{Morsink} \& {Stella}(1999)}]{most1999}
{Morsink}, S.~M. \& {Stella}, L. 1999, \apj, 513, 827

\bibitem[{{Osherovich} \& {Titarchuk}(1999)}]{osti1999}
{Osherovich}, V. \& {Titarchuk}, L. 1999, \apjl, 523, L73

\bibitem[{{Penninx} {et~al.}(1988){Penninx}, {Lewin}, {Zijlstra}, {Mitsuda}, \&
  {van Paradijs}}]{pelezi1988}
{Penninx}, W., {Lewin}, W. H.~G., {Zijlstra}, A.~A., {Mitsuda}, K., \& {van
  Paradijs}, J. 1988, \nat, 336, 146

\bibitem[{{Press} {et~al.}(1992){Press}, {Teukolsky}, {Vetterling}, \&
  {Flannery}}]{prteve1992}
{Press}, W.~H., {Teukolsky}, S.~A., {Vetterling}, W.~T., \& {Flannery}, B.~P.
  1992, Numerical recipes in FORTRAN. The art of scientific computing
  (Cambridge: University Press, |c1992, 2nd ed.)

\bibitem[{{Priedhorsky} {et~al.}(1986){Priedhorsky}, {Hasinger}, {Lewin},
  {Middleditch}, {Parmar}, {Stella}, \& {White}}]{prhale1986}
{Priedhorsky}, W., {Hasinger}, G., {Lewin}, W. H.~G., {et~al.} 1986, \apjl,
  306, L91

\bibitem[{{Psaltis} {et~al.}(1999){Psaltis}, {Belloni}, \& {van der
  Klis}}]{psbeva1999}
{Psaltis}, D., {Belloni}, T., \& {van der Klis}, M. 1999, \apj, 520, 262

\bibitem[{{Psaltis} {et~al.}(1995){Psaltis}, {Lamb}, \& {Miller}}]{pslami1995}
{Psaltis}, D., {Lamb}, F.~K., \& {Miller}, G.~S. 1995, \apjl, 454, L137

\bibitem[{{Psaltis} {et~al.}(1998){Psaltis}, {M\'endez}, {Wij\-nands}, {Homan},
  {Jonker}, {van der Klis}, {Lamb}, {Kuul\-kers}, {van Paradijs}, \&
  {Lewin}}]{psmewi1998}
{Psaltis}, D., {M\'endez}, M., {Wij\-nands}, R., {et~al.} 1998, \apjl, 501, L95

\bibitem[{{Psaltis} \& {Norman}(2001)}]{psno2001}
{Psaltis}, D. \& {Norman}, C. 2001, \apj, submitted

\bibitem[{{Rutledge} {et~al.}(1999){Rutledge}, {Lewin}, {van der Klis}, {van
  Paradijs}, {Dotani}, {Vaughan}, {Belloni}, {Oosterbroek}, \&
  {Kouveliotou}}]{ruleva1999}
{Rutledge}, R.~E., {Lewin}, W. H.~G., {van der Klis}, M., {et~al.} 1999, \apjs,
  124, 265

\bibitem[{{Sobczak} {et~al.}(2000){Sobczak}, {McClintock}, {Remillard}, {Cui},
  {Levine}, {Morgan}, {Orosz}, \& {Bailyn}}]{somcre2000}
{Sobczak}, G.~J., {McClintock}, J.~E., {Remillard}, R.~A., {et~al.} 2000, \apj,
  531, 537

\bibitem[{{Stella} \& {Vietri}(1998)}]{stvi1998}
{Stella}, L. \& {Vietri}, M. 1998, \apjl, 492, L59

\bibitem[{{Stella} \& {Vietri}(1999)}]{stvi1999}
---. 1999, \prl, 82, 17

\bibitem[{{Stella} {et~al.}(1999){Stella}, {Vietri}, \& {Morsink}}]{stvimo1999}
{Stella}, L., {Vietri}, M., \& {Morsink}, S.~M. 1999, \apjl, 524, L63

\bibitem[{{Strickman} \& {Barret}(2000)}]{stba2000}
{Strickman}, M. \& {Barret}, D. 2000, in Proceedings of the fifth Compton
  Symposium, held in Portsmouth, NH, USA, September 1999. Melville, NY:
  American Institute of Physics (AIP), 2000. Edited by Mark L. McConnell and
  James M. Ryan AIP Conference Proceedings, Vol. 510., p. 157

\bibitem[{{Tan} {et~al.}(1992){Tan}, {Lewin}, {Hjellming}, {Penninx}, {van
  Paradijs}, {van der Klis}, \& {Mitsuda}}]{talehj1992}
{Tan}, J., {Lewin}, W. H.~G., {Hjellming}, R.~M., {et~al.} 1992, \apj, 385, 314

\bibitem[{{Tanaka} \& {Lewin}(1995)}]{tale1995}
{Tanaka}, Y. \& {Lewin}, W. H.~G. 1995, in X-ray binaries (Cambridge
  Astrophysics Series, Cambridge, MA: Cambridge University Press, |c1995,
  edited by Lewin, Walter H.G.; Van Paradijs, Jan; Van den Heuvel, Edward
  P.J.), p. 126

\bibitem[{{van der Klis}(1986)}]{va1986}
{van der Klis}, M. 1986, in LNP Vol. 266: The Physics of Accretion onto Compact
  Objects, p. 157

\bibitem[{{van der Klis}(1989)}]{va1989}
{van der Klis}, M. 1989, in Proceedings of the NATO Advanced Study Institute on
  Timing Neutron Stars, held in \c Ce\c sme, \. Izmir, Turkey, April 4--15,
  1988. Editors, H. \"Ogelman and E.P.J. van den Heuvel; Publisher, Kluwer
  Academic, Dordrecht, The Netherlands, Boston, Massachusetts, p. 27

\bibitem[{{van der Klis}(1991)}]{va1991}
{van der Klis}, M. 1991, in NATO ASIC Proc. 344: Neutron Stars, p. 319

\bibitem[{{van der Klis}(1995{\natexlab{a}})}]{va1995a}
{van der Klis}, M. 1995{\natexlab{a}}, in X-ray binaries (Cambridge
  Astrophysics Series, Cambridge, MA: Cambridge University Press, |c1995,
  edited by Lewin, Walter H.G.; Van Paradijs, Jan; Van den Heuvel, Edward
  P.J.), p. 252

\bibitem[{{van der Klis}(1995{\natexlab{b}})}]{va1995b}
{van der Klis}, M. 1995{\natexlab{b}}, in Proceedings of the NATO Advanced
  Study Institute on the Lives of the Neutron Stars, held in Kemer, Turkey,
  August 29-September 12, 1993. Editor(s), M.A. Alpar, U. Kiziloglu, J. van
  Paradijs; Publisher, Kluwer Academic, Dordrecht, The Netherlands, Boston,
  Massachusetts, 1995., p. 301

\bibitem[{{van der Klis}(2000)}]{va2000}
---. 2000, \araa, 38, 717

\bibitem[{{van der Klis} {et~al.}(1997{\natexlab{a}}){van der Klis}, {Homan},
  {Wij\-nands}, {Kuul\-kers}, {Lamb}, {Psaltis}, {Dieters}, {van Paradijs},
  {Lewin}, \& {Vaughan}}]{vahowi1997}
{van der Klis}, M., {Homan}, J., {Wij\-nands}, R., {et~al.} 1997{\natexlab{a}},
  \iaucirc, 6565

\bibitem[{{van der Klis} {et~al.}(1996){van der Klis}, {Swank}, {Zhang},
  {Jahoda}, {Morgan}, {Lewin}, {Vaughan}, \& {van Paradijs}}]{vaswzh1996}
{van der Klis}, M., {Swank}, J.~H., {Zhang}, W., {et~al.} 1996, \apjl, 469, L1

\bibitem[{{van der Klis} {et~al.}(1997{\natexlab{b}}){van der Klis},
  {Wij\-nands}, {Horne}, \& {Chen}}]{vawiho1997}
{van der Klis}, M., {Wij\-nands}, R. A.~D., {Horne}, K., \& {Chen}, W.
  1997{\natexlab{b}}, \apjl, 481, L97

\bibitem[{{van Paradijs} {et~al.}(1990){van Paradijs}, {Allington-Smith},
  {Callanan}, {Charles}, {Hassall}, {Machin}, {Mason}, {Naylor}, \&
  {Smale}}]{vaalca1990}
{van Paradijs}, J., {Allington-Smith}, J., {Callanan}, P., {et~al.} 1990, \aap,
  235, 156

\bibitem[{{Vrtilek} {et~al.}(1991){Vrtilek}, {Penninx}, {Raymond}, {Verbunt},
  {Hertz}, {Wood}, {Lewin}, \& {Mitsuda}}]{vrpera1991}
{Vrtilek}, S.~D., {Penninx}, W., {Raymond}, J.~C., {et~al.} 1991, \apj, 376,
  278

\bibitem[{{Vrtilek} {et~al.}(1990){Vrtilek}, {Raymond}, {Garcia}, {Verbunt},
  {Hasinger}, \& {Kurster}}]{vrraga1990}
{Vrtilek}, S.~D., {Raymond}, J.~C., {Garcia}, M.~R., {et~al.} 1990, \aap, 235,
  162

\bibitem[{{Wij\-nands} {et~al.}(1998{\natexlab{a}}){Wij\-nands}, {Homan}, {van
  der Klis}, {Kuul\-kers}, {van Paradijs}, {Lewin}, {Lamb}, {Psaltis}, \&
  {Vaughan}}]{wihova1998}
{Wij\-nands}, R., {Homan}, J., {van der Klis}, M., {et~al.} 1998{\natexlab{a}},
  \apjl, 493, L87

\bibitem[{{Wij\-nands} {et~al.}(1997{\natexlab{a}}){Wij\-nands}, {Homan}, {van
  der Klis}, {M\'endez}, {Kuul\-kers}, {van Paradijs}, {Lewin}, {Lamb},
  {Psaltis}, \& {Vaughan}}]{wihova1997}
---. 1997{\natexlab{a}}, \apjl, 490, L157

\bibitem[{{Wij\-nands} {et~al.}(1998{\natexlab{b}}){Wij\-nands}, {M\'endez},
  {van der Klis}, {Psaltis}, {Kuul\-kers}, \& {Lamb}}]{wimeva1998}
{Wij\-nands}, R., {M\'endez}, M., {van der Klis}, M., {et~al.}
  1998{\natexlab{b}}, \apjl, 504, L35

\bibitem[{{Wij\-nands} \& {van der Klis}(1999{\natexlab{a}})}]{wiva1999a}
{Wij\-nands}, R. \& {van der Klis}, M. 1999{\natexlab{a}}, \apj, 514, 939

\bibitem[{{Wij\-nands} \& {van der Klis}(1999{\natexlab{b}})}]{wiva1999b}
---. 1999{\natexlab{b}}, \apj, 522, 965

\bibitem[{{Wij\-nands} \& {van der Klis}(2001)}]{wiva2001}
---. 2001, \mnras, in press

\bibitem[{{Wij\-nands} {et~al.}(1997{\natexlab{b}}){Wij\-nands}, {van der
  Klis}, {Kuul\-kers}, {Asai}, \& {Hasinger}}]{wivaku1997}
{Wij\-nands}, R., {van der Klis}, M., {Kuul\-kers}, E., {Asai}, K., \&
  {Hasinger}, G. 1997{\natexlab{b}}, \aap, 323, 399

\bibitem[{{Wij\-nands} {et~al.}(1996){Wij\-nands}, {van der Klis}, {Psaltis},
  {Lamb}, {Kuul\-kers}, {Dieters}, {van Paradijs}, \& {Lewin}}]{wivaps1996}
{Wij\-nands}, R., {van der Klis}, M., {Psaltis}, D., {et~al.} 1996, \apjl, 469,
  L5

\bibitem[{{Wij\-nands} {et~al.}(1999){Wij\-nands}, {van der Klis}, \&
  {Rijkhorst}}]{wivari1999}
{Wij\-nands}, R., {van der Klis}, M., \& {Rijkhorst}, E. 1999, \apjl, 512, L39

\bibitem[{{Yu} {et~al.}(2001){Yu}, {van der Klis}, \& {Jonker}}]{yuvajo2001}
{Yu}, W., {van der Klis}, M., \& {Jonker}, P. 2001, \apj, in prep.

\bibitem[{{Zhang}(1995)}]{zh1995}
{Zhang}, W. 1995, XTE/PCA Internal Memo

\bibitem[{{Zhang} {et~al.}(1995){Zhang}, {Jahoda}, {Swank}, {Morgan}, \&
  {Giles}}]{zhjasw1995}
{Zhang}, W., {Jahoda}, K., {Swank}, J.~H., {Morgan}, E.~H., \& {Giles}, A.~B.
  1995, \apj, 449, 930

\bibitem[{{Zhang} {et~al.}(1998){Zhang}, {Strohmayer}, \& {Swank}}]{zhstsw1998}
{Zhang}, W., {Strohmayer}, T.~E., \& {Swank}, J.~H. 1998, \apjl, 500, L167

\end{thebibliography}



\newpage
\clearpage

\newpage
\clearpage

\newpage
\clearpage

\newpage
\clearpage

\newpage
\clearpage

\newpage
\clearpage

\newpage
\clearpage

\newpage
\clearpage

\newpage
\clearpage

\newpage
\clearpage

\newpage
\clearpage

\newpage
\clearpage

\newpage
\clearpage

\newpage
\clearpage

\newpage
\clearpage

\begin{table}[t]
\begin{center}
\begin{tabular}{llccc}
\hline \hline
Begin Time (UTC) & End Time (UTC) & Total (ks) & Modes$^a$ & Gain Epoch \\
\hline
1997-02-02 19:13 & 1997-02-27 03:34  & 58.7  & 3 or 4,5,6,9 & 3 \\
1997-04-01 19:13 & 1997-04-04 23:26  & 34.6  & 4,5,6,9      & 3 \\
1997-07-27 02:13 & 1997-07-28 00:33  & 42.9  & 4,5,6,9      & 3 \\
1998-08-07 06:40 & 1998-08-08 23:40  & 71.0  & 4,5,8        & 3 \\
1998-11-18 06:42 & 1998-11-20 13:31  & 86.0  & 4,5(,7),8    & 3 \\
1999-10-03 02:43 & 1999-10-12 07:05  & 297.6 & 10,11,12     & 4 \\
2000-03-31 12:15 & 2000-03-31 16:31  & 6.9   & 10,11,12     & 4 \\            
\hline
\end{tabular}
\\ \vspace{0.2cm} $^a$ Modes in addition to the Standard 1 and
Standard 2 modes. See Table \ref{gx17+2_modes_tab} for modes.

\caption{ A log of all \xte/PCA observations used in this
paper. Mode 7 was not always active during the November 1998
observations. Note that none of the observations represents an
uninterrupted interval. Each is a collection of observations that
were done around the same time. These observations were separated in
time from each other for various reasons, such as Earth occultations,
passages of the South Atlantic Anomaly, or observations of
other sources.\label{obs_tab}} \end{center} 
\end{table}

\newpage
\clearpage

\begin{table}[t]
\begin{center}
\begin{tabular}{clccc}
\hline \hline
Mode & Name & Time Resolution (s) & Energy range (keV) & Energy channels \\
1 & Standard 1 & $2^{-3}$  & 2--60 & 1 \\
2 & Standard 2 & $2^4$     & 2--60 & 129 \\
\hline
3 & E\_8us\_8A\_0\_1s      & \raisebox{.0ex}[2.5ex][.0ex]{$2^{-17}$}  & 2--60    &  8 \\ 
4 & SB\_125us\_0\_13\_1s   & $2^{-13}$  & 2--5.1   &  1 \\ 
5 & SB\_125us\_14\_17\_1s  & $2^{-13}$  & 5.1--6.6 &  1 \\ 
6 & SB\_125us\_18\_23\_1s  & $2^{-13}$  & 6.6--8.7 &  1 \\ 
7 & SB\_125us\_18\_249\_1s & $2^{-13}$  & 6.6--60  &  1 \\ 
8 & E\_16us\_64M\_18\_1s   & $2^{-16}$  & 6.6--60  &  64 \\ 
9 & SB\_125us\_24\_249\_1s & $2^{-13}$  & 8.7--60  &  1 \\ 
\hline
10 & SB\_125us\_0\_13\_1s   & \raisebox{.0ex}[2.5ex][.0ex]{$2^{-13}$}  & 2--5.8   &  1 \\ 
11 & SB\_125us\_14\_17\_1s  & $2^{-13}$  & 5.8--7.5 &  1 \\ 
12 & E\_16us\_64M\_18\_1s   & $2^{-16}$  & 7.5--60  &  64 \\ 
\hline 
\end{tabular}

\caption{Names and settings of the data modes that were
used in our analysis. The lower and upper energy boundaries of the
PCA energy sensitivity range are given as 2 and 60 keV, although they
changed between (and also slightly during) the different gain
epochs.\label{gx17+2_modes_tab}} \end{center} 
\end{table}

\newpage
\clearpage

\begin{table}[t]
\begin{center}
{\footnotesize
\begin{tabular}{ccccccc}
\hline 
\hline
\multicolumn{1}{c}{Gain Epoch} & \multicolumn{2}{c}{Soft Color} & \multicolumn{2}{c}{Hard Color} & \multicolumn{2}{c}{Intensity}  \\
\hline
                 & (Channels)       & (keV)                 & (Channels)        & (keV)        & (Channels)        & (keV)\\
3                & 10--16/5--9  & 4.8--7.3/3.0--4.8 & 26--50/17--25 & 10.5--19.7/7.3--10.5     & 5--50             & 3.0--19.7 \\
4                & 8--13/4--7   & 4.6--7.1/2.9--4.6 & 22--42/14--21 & 10.5--19.6/7.1--10.5     & 4--42             & 2.9--19.6 \\
\hline
\end{tabular}
}
\caption{ Channel and energy boundaries of the soft and hard colors
and the intensity used for the spectral analysis. The channel numbers
refer to the Standard 2 mode channels (1--129).\label{colors_tab}}
\end{center} 
\end{table}

\newpage
\clearpage

\begin{table}[t]
\begin{center}
\begin{tabular}{lc}
\hline \hline Component & \sz-range \\ 
\hline 
LFN & $-$0.6--5.0 \\ 
HBO (fund.) & $-$0.6--2.1\\ 
HBO (2nd harm.) & $-$0.6--1.0\\ 
HBO (sub-harm.) & $-$0.6--1.3 \\ 
Upper kHz QPO & $-$0.3--1.7 \\ 
VLFN & 0.3--5.5\\ 
Lower kHz QPO & 0.5--1.5 \\ 
NBO & 1.6--2.1\\ 
FBO & 2.0--2.7\\
\hline
\end{tabular}
\caption{The nine different components in the combined
epoch 3/epoch 4 power spectra, and the \sz-ranges in which they were
detected. The components are listed in order of
\sz\ appearance.\label{sz-comp_tab}} \end{center} 
\end{table}

\newpage
\clearpage

\pagestyle{empty}

\begin{sidewaystable}[b]
\renewcommand{\arraystretch}{0.8}
{\scriptsize

\begin{tabular}{c@{\extracolsep{4pt}}c@{\extracolsep{2pt}}c@{\extracolsep{2pt}}c@{\extracolsep{2pt}}c@{\extracolsep{2pt}}c@{\extracolsep{2pt}}c@{\extracolsep{2pt}}c@{\extracolsep{2pt}}c@{\extracolsep{2pt}}c@{\extracolsep{0pt}}c@{\extracolsep{2pt}}c@{\extracolsep{2pt}}c}
\hline
\hline
                &  \multicolumn{3}{c}{HBO Fundamental}                                         &  \multicolumn{3}{c}{HBO 2$^{nd}$ Harmonic}				&  \multicolumn{3}{c}{HBO Sub-harmonic} 				     & \multicolumn{3}{c}{NBO} \\						 	\\
$S_z$           &  Rms$^a$ (\%)               & FWHM (Hz)               & Frequency (Hz)       &  Rms$^a$ (\%)		  & FWHM (Hz)		   & Frequency (Hz)	&  Rms$^b$ (\%)		   & FWHM (Hz)  	    & Frequency (Hz)	     &  Rms$^b$ (\%)	      & FWHM (Hz)		& Frequency (Hz)	 	\\
\hline
$-0.43\pm$0.05  &  4.3$\pm$0.7  	  & 6$^{+4}_{-2}$	    &  21.3$\pm$0.8	       &  4.5$\pm$0.7		  &  9$^{+4}_{-3}$	   &  41.4$\pm$1.4	&  6.4$\pm$1.0		   &  27$^{+15}_{-10}$	    &  0 {\it (fixed)}       &  \ldots  	      &  \ldots 		&  \ldots			\\
$-0.32\pm$0.04  &  3.9$\pm$0.6  	  & 6.9$^{+2.8}_{-1.9}$     &  21.4$\pm$0.4	       &  4.9$\pm$0.3		  &  11.7$\pm$1.7	   &  41.4$\pm$0.5	&  6.9$\pm$0.4		   &  19$\pm$6		    &  0 {\it (fixed)}       &  \ldots  	      &  \ldots 		&  \ldots			\\
$-0.23\pm$0.03  &  2.9$\pm$0.3  	  & 4.1$\pm$0.9 	    &  22.2$\pm$0.2	       &  3.9$\pm$0.2		  &  7.8$\pm$1.1	   &  43.9$\pm$0.3	&  7.4$\pm$0.3 		   &  44$\pm$6		    &  0 {\it (fixed)}       &  \ldots  	      &  \ldots 		&  \ldots			\\
$-0.15\pm$0.03  &  3.8$\pm$0.3  	  & 7.3$\pm$1.0 	    &  22.6$\pm$0.3	       &  5.14$^{+0.19}_{-0.30}$  &  15.8$\pm$1.4	   &  43.9$\pm$0.4	&  6.7$\pm$0.2 		   &  24$^{+8}_{-5}$	    &  0 {\it (fixed)}       &  \ldots  	      &  \ldots 		&  \ldots			\\
$-0.04\pm$0.03  &  4.02$\pm$0.20	  & 8.7$\pm$0.9 	    &  24.3$\pm$0.2	       &  4.79$\pm$0.12 	  &  15.0$\pm$1.0	   &  47.3$\pm$0.3	&  6.41$\pm$0.17	   &  19$^{+4}_{-3}$	    &  0 {\it (fixed)}       &  \ldots  	      &  \ldots 		&  \ldots			\\
0.05$\pm$0.03   &  3.12$\pm$0.14          & 5.4$\pm$0.5 	    &  26.39$\pm$0.13	       &  4.06$\pm$0.14 	  &  12.7$\pm$0.9	   &  50.7$\pm$0.2	&  7.0$\pm$0.9		   &  29$\pm$3		    &  0$\pm$4  	     &  \ldots  	      &  \ldots 		&  \ldots			\\
0.16$\pm$0.03   &  3.21$\pm$0.18          & 6.1$\pm$0.6 	    &  27.42$\pm$0.14 	       &  3.82$\pm$0.13 	  &  12.4$\pm$0.9	   &  53.7$\pm$0.3	&  6.6$^{+1.7}_{-0.9}$     &  31$\pm$4		    &  2 $^{+3}_{-6}$	     &  \ldots  	      &  \ldots 		&  \ldots			\\
0.25$\pm$0.03   &  3.54$\pm$0.17          & 8.8$\pm$0.7 	    &  30.20$\pm$0.18 	       &  3.73$\pm$0.12 	  &  16.4$\pm$1.2	   &  57.9$\pm$0.4	&  5.7$\pm$0.8		   &  25$\pm$3		    &  8$\pm$3  	     &  \ldots  	      &  \ldots 		&  \ldots			\\
0.35$\pm$0.04   &  2.81$\pm$0.11          & 6.8$\pm$0.5 	    &  33.33$\pm$0.15	       &  2.86$\pm$0.12 	  &  15.0$\pm$1.3	   &  63.6$\pm$0.4	&  5.2$^{+1.7}_{-0.5}$     &  32$^{+5}_{-3}$	    &  9$^{+2}_{-8}$	     &  \ldots  	      &  \ldots 		&  \ldots			\\
0.46$\pm$0.04   &  2.95$\pm$0.11          & 8.3$\pm$0.5 	    &  35.35$\pm$0.15	       &  2.61$\pm$0.09 	  &  13.9$\pm$1.2	   &  68.5$\pm$0.4	&  4.6$^{+0.9}_{-0.4}$     &  28$\pm$3		    &  10$^{+1.8}_{-4.2}$    &  \ldots  	      &  \ldots 		&  \ldots			\\
0.56$\pm$0.03   &  2.75$\pm$0.09          & 8.3$\pm$0.5 	    &  37.91$\pm$0.14	       &  2.41$\pm$0.08 	  &  13.6$\pm$1.1	   &  73.4 $\pm$0.3	&  4.1$\pm$0.4		   &  26$\pm$3  	    &  15.0$\pm$1.5	     &  \ldots  	      &  \ldots 		&  \ldots			\\
0.65$\pm$0.03   &  2.57$\pm$0.10          & 7.8$\pm$0.6 	    &  39.56$\pm$0.14 	       &  2.42$\pm$0.09 	  &  16.8$\pm$1.5	   &  77.9$\pm$0.5	&  3.9$\pm$0.2 		   &  28$\pm$3  	    &  18.0$\pm$1.0	     &  \ldots  	      &  \ldots 		&  \ldots			\\
0.75$\pm$0.03   &  2.41$\pm$0.13          & 9.1$\pm$0.9 	    &  42.9$\pm$0.2 	       &  1.92$\pm$0.11 	  &  14.5$\pm$1.9	   &  83.3$\pm$0.7	&  3.6$\pm$0.3		   &  28$\pm$4  	    &  20.6$\pm$1.0	     &  \ldots  	      &  \ldots 		&  \ldots			\\
0.85$\pm$0.03   &  2.2$\pm$0.2            & 9.8$\pm$1.6 	    &  45.3$\pm$0.4	       &  1.62$\pm$0.15 	  &  13$\pm$3		   &  89.4$\pm$0.9	&  3.2$\pm$0.3 		   &  29$\pm$5  	    &  24.9$\pm$1.4	     &  \ldots  	      &  \ldots 		&  \ldots			\\
0.95$\pm$0.03   &  2.3$\pm$0.3            & 13.$\pm$3		    &  48.9$\pm$0.7 	       &  1.8$\pm$0.2		  &  22$^{+8}_{-5 }$	   &  93$\pm$2  	&  2.5$\pm$0.4 		   &  21$\pm$7  	    &  25.2$\pm$1.3	     &  \ldots  	      &  \ldots 		&  \ldots			\\
1.06$\pm$0.03   &  1.52$\pm$0.17          & 6.0$\pm$1.3 	    &  53.9$\pm$0.4	       &  1.09$\pm$0.15 	  &  15 {\it (fixed)}	   &  107 {\it (fixed)} &  2.76$^{+0.26}_{-0.19}$  &  31$\pm$7	            &  32.2$\pm$1.6	     &  \ldots		      &  \ldots		        &  \ldots			  \\
1.15$\pm$0.03   &  1.40$\pm$0.12          & 5.1$\pm$1.0 	    &  56.7$\pm$0.3 	       &  $<$0.7		  &  15 {\it (fixed)}	   &  113 {\it (fixed)} &  2.5$^{+0.3}_{-0.2}$     &  32$\pm$6	            &  35.5$\pm$1.6	     &  \ldots		      &  \ldots		        &  \ldots			  \\
1.25$\pm$0.03   &  1.42$\pm$0.12          & 5.1$^{+0.6}_{-0.3}$     &  58.5$\pm$0.3	       &  \ldots		  & \ldots		   &  \ldots		&  2.8$^{+0.5}_{-0.3}$     &  57$^{+17}_{-12}$      &  41$\pm$6	             &  \ldots		      &  \ldots		        &  \ldots			  \\
1.35$\pm$0.03   &  1.63$\pm$0.11          & 4.0$\pm$0.6 	    &  60.11$\pm$0.17	       &  \ldots		  & \ldots		   &  \ldots		&  $<2.5$	           &  21$\pm$14		    &  43$\pm$3		     &  \ldots		      &  \ldots		        &  \ldots 		      \\
1.45$\pm$0.04   &  1.40$\pm$0.12          & 5.2$\pm$1.2 	    &  60.3$\pm$0.3 	       &  \ldots		  & \ldots		   &  \ldots		&  \ldots		   &  \ldots		    &  \ldots		     &  \ldots  	      &  \ldots 		&  \ldots		 	\\
1.55$\pm$0.03   &  1.8$\pm$0.2            & 14$\pm$3		    &  56.7$^{+0.5}_{-0.8}$    &  \ldots		  & \ldots		   &  \ldots		&  \ldots		   &  \ldots		    &  \ldots		     &  \ldots  	      &  \ldots 		&  \ldots		 	\\
1.65$\pm$0.04   &  1.72$\pm$0.14          & 12$\pm$3		    &  54.6$\pm$1.0            &  \ldots		  & \ldots		   &  \ldots		&  \ldots		   &  \ldots		    &  \ldots		     &  1.7$\pm$0.2 	     &  3.8$\pm$0.7	       &  6.33$\pm$0.16 	       \\
1.76$\pm$0.04   &  1.99$\pm$0.12          & 17$\pm$3		    &  53.6$\pm$1.0            &  \ldots		  & \ldots		   &  \ldots		&  \ldots		   &  \ldots		    &  \ldots		     &  2.07$\pm$0.11	     &  2.08$\pm$0.20	       &  6.82$\pm$0.04 	       \\
1.86$\pm$0.04   &  1.90$\pm$0.12          & 17$\pm$3		    &  52.6$\pm$1.1            &  \ldots		  & \ldots		   &  \ldots		&  \ldots		   &  \ldots		    &  \ldots		     &  2.89$\pm$0.08 	     &  2.13$\pm$0.11	       &  6.91$\pm$0.03 	       \\
1.95$\pm$0.04   &  1.80$\pm$0.16          & 19$\pm$4		    &  50.1$\pm$1.2            &  \ldots		  & \ldots		   &  \ldots		&  \ldots		   &  \ldots		    &  \ldots		     &  2.94$\pm$0.12	     &  2.11$\pm$0.13	       &  6.96$\pm$0.02 	       \\
2.04$\pm$0.04   &  1.2$\pm$0.2            & 11$^{+7}_{-5}$          &  48.2$\pm$2	       &  \ldots		  & \ldots		   &  \ldots		&  \ldots		   &  \ldots		    &  \ldots		     &  3.2$\pm$0.2 	     &  3.7$\pm$0.4	       &  6.86$\pm$0.10 	       \\
                & \ldots		  & \ldots		    &  \ldots		       &  \ldots		  & \ldots		   &  \ldots		&  \ldots		   &  \ldots		    &  \ldots		     &  2.8$\pm$0.8 	     &  5.4$^{+0.6}_{-1.4}$    &  13.9$^{+0.6}_{-0.4}$         \\
2.13$\pm$0.05   &  $<$1.2		  & 9 {\it (fixed)}	    &  46 {\it (fixed)}	       &  \ldots		  & \ldots		   &  \ldots		&  \ldots		   &  \ldots		    &  \ldots		     &  4.93$\pm$0.11 	     &  5.1$\pm$0.2	       &  14.78$\pm$0.08	       \\
2.23$\pm$0.06   & \ldots		  & \ldots		    &  \ldots		       &  \ldots		  & \ldots		   &  \ldots		&  \ldots		   &  \ldots		    &  \ldots		     &  5.60$\pm$0.07	     &  6.4$\pm$0.18	       &  15.44$\pm$0.06	       \\
2.34$\pm$0.05   & \ldots		  & \ldots		    &  \ldots		       &  \ldots		  & \ldots		   &  \ldots		&  \ldots		   &  \ldots		    &  \ldots		     &  5.28$\pm$0.11 	     &  8.1$\pm$0.3	       &  16.75$\pm$0.12	       \\
2.45$\pm$0.05   & \ldots		  & \ldots		    &  \ldots		       &  \ldots		  & \ldots		   &  \ldots		&  \ldots		   &  \ldots		    &  \ldots		     &  4.37$\pm$0.17 	     &  10.3$\pm$0.7	       &  18.2$\pm$0.2  	       \\
2.54$\pm$0.04   & \ldots		  & \ldots		    &  \ldots		       &  \ldots		  & \ldots		   &  \ldots		&  \ldots		   &  \ldots		    &  \ldots		     &  3.36$\pm$0.20	     &  13.6$\pm$1.6	       &  19.7$\pm$0.6  	       \\
2.65$\pm$0.04   & \ldots		  & \ldots		    &  \ldots		       &  \ldots		  & \ldots		   &  \ldots		&  \ldots		   &  \ldots		    &  \ldots		     &  1.8$\pm$0.4 	     &  8$\pm$4 	       &  23.1$\pm$0.9  	       \\
2.75$\pm$0.04   & \ldots		  & \ldots		    &  \ldots		       &  \ldots		  & \ldots		   &  \ldots		&  \ldots		   &  \ldots		    &  \ldots		     &  $<1.8$	             &  10 {\it (fixed)}       &  25 {\it (fixed)}	       \\
\hline
\end{tabular}
\\
\noindent $^a$ Integrated between $-\infty$ Hz and +$\infty$ Hz. \\
\noindent $^b$ Integrated between 0 Hz and +$\infty$ Hz, because of low Q values. 
} 
\caption{Fit results for the low frequency QPOs.  \label{low_tab}} 
\end{sidewaystable}

\newpage
\clearpage

\pagestyle{plain}

\begin{table}[t]
\begin{center}
\begin{tabular}{lccc}
\hline
\hline
Interval & Count Rate        & Fractional rms & Absolute rms  \\
         & (counts s$^{-1}$) &   (\%)         & (counts s$^{-1}$) \\
\hline
Nov 1998 - I   & 2811 & 2.58$\pm$0.19 & 72.5$\pm$5.3 \\
Nov 1998 - II  & 2648 & 3.2$\pm$0.3   & 85$\pm$8 \\
Nov 1998 - III & 4568 & $<$0.71       & $<$32.4 \\
Nov 1998 - IV  & 3232 & $<$2.1        & $<$68 \\
Nov 1998 - V   & 2833 & 3.3$\pm$0.2   & 93$\pm$6 \\
\hline
Oct 1999 - I   & 1742  & 5.56$\pm$.17  & 96.9$\pm$3.0\\
Oct 1999 - II  & 2616 & $<$1.57       & $<$41.1 \\
Oct 1999 - III & 1847 & 4.3$\pm$0.4   & 79$\pm$7\\
\hline
\end{tabular}
\caption{The properties of the NBO/FBO during the two bursts
shown in Figure \ref{burst-nbo_fig}. The intervals given in first
column can also be found in that figure. The upper limits in the last
two columns are 95\% confidence. The count rates are in the 5.1--60 keV (November 1998, 5 PCUs) and 5.8--60 keV (October 1999, 3 PCUs) bands.\label{burst-nbo_tab}} \end{center}
\end{table}

\newpage
\clearpage

\pagestyle{empty}

\renewcommand{\arraystretch}{0.9}

\begin{table}[t]
{\scriptsize
\begin{tabular}{ccccccccc}
\hline
\hline
$S_z$           &         \multicolumn{2}{c}{VLFN}	                      &        	\multicolumn{3}{c}{LFN (Cut-off power law)}                                                         &		\multicolumn{3}{c}{LFN (Lorentzian)}  									\\
                &         Rms$^a$ (\%)         &        Index                     &        Rms$^b$ (\%)                     &     Index                      &        Cut-off  (Hz)         &             Rms$^b$ (\%)               &        FWHM (Hz)            &        Frequency (Hz)                      \\ 
\hline
$-0.43\pm$0.05  &	  \ldots	   &       \ldots	     	      &        \ldots 	                   &	    \ldots		    &	     \ldots		            &	          2.5$\pm$0.5 		     &        1.4$\pm$0.7	   &	    0.9$\pm$0.3 			\\
$-0.32\pm$0.04  &	  \ldots	   &       \ldots	     	      &        \ldots 	     	           &	    \ldots		    &	     \ldots		            &	          3.7$\pm$0.9 		     &        3.1$\pm$1.0	   &	    1.05$\pm$0.19			\\
$-0.23\pm$0.03  &	  \ldots	   &       \ldots	     	      &        \ldots 	    	           &	    \ldots		    &	     \ldots		            &	          5.0$\pm$0.3 		     &        4.0$\pm$0.4	   &	    1.15$\pm$0.13			\\
$-40.15\pm$0.03  &	  \ldots	   &       \ldots	     	      &        \ldots 	    	           &	    \ldots		    &	     \ldots		            &	          4.5$\pm$0.5 		     &        3.9$\pm$0.6	   &	    1.19$\pm$0.09			\\
$-0.04\pm$0.03  &	  \ldots	   &       \ldots	     	      &        \ldots 	    	           &	    \ldots		    &	     \ldots		            &	          4.2$\pm$0.4 		     &        3.4$\pm$0.4	   &	    1.30$\pm$0.06			\\
0.05$\pm$0.03	&	  \ldots	   &       \ldots	     	      &        4.1$\pm$0.3                 &	$-0.76\pm$0.11		    &         1.43$\pm$0.15		    &		  \ldots        	     &        \ldots		   &	    \ldots				\\
0.16$\pm$0.03	&	 $<$0.9            &	 0.6 {\it (fixed)}            &        4.35$\pm$0.15	           &	$-0.80\pm$0.11  	    &	      1.56$^{+0.16}_{-0.11}$        &		  \ldots        	     &        \ldots		   &	    \ldots				\\
0.25$\pm$0.03	&	 $<$0.7	           &	 0.6 {\it (fixed)}	      &        4.6$\pm$0.3	           &	$-0.64\pm$0.07  	    &	      1.93$\pm$0.19		    &		  \ldots        	     &        \ldots		   &	    \ldots				\\
0.35$\pm$0.04	&	 0.86$\pm$0.07     &	 0.69$\pm$0.04  	      &        4.46$^{+0.14}_{-0.24}$      &	$-0.86\pm$0.07  	    &	      1.85$^{+0.12}_{-0.08}$  	    &		  \ldots        	     &        \ldots		   &	    \ldots				\\
0.46$\pm$0.04	&	 0.83$\pm$0.06     &	 0.61$\pm$0.03  	      &        4.44$^{+0.13}_{-0.23}$      &	$-0.86^{+0.03}_{-0.06}$     &         2.01$^{+0.06}_{-0.12}$  	    &		  \ldots        	     &        \ldots		   &	    \ldots				\\
0.56$\pm$0.03	&	 0.65$\pm$0.06     &	 0.66$\pm$0.05  	      &        4.66$\pm$0.11	           &	$-0.71\pm$0.05  	    &	      2.54$\pm$0.14 		    &		  \ldots        	     &        \ldots		   &	    \ldots				\\
0.65$\pm$0.03	&	 0.65$\pm$0.06     &	 0.85$^{+0.15}_{-0.09}$       &        4.74$\pm$0.07	           &	$-0.67\pm$0.04  	    &	      2.94$\pm$0.12 		    &		  \ldots        	     &        \ldots		   &	    \ldots				\\
0.75$\pm$0.03	&	 0.74$\pm$0.05     &	 0.57$\pm$0.03  	      &        4.55$\pm$0.07	           &	$-0.70\pm$0.05  	    &	      3.24$\pm$0.14		    &		  \ldots        	     &        \ldots		   &	    \ldots				\\
0.85$\pm$0.03	&	 0.61$\pm$0.08     &	 0.47$\pm$0.05  	      &        4.53$\pm$0.08	           &	$-0.61\pm$0.06  	    &	      4.0$\pm$0.2 		    &		  \ldots        	     &        \ldots		   &	    \ldots				\\
0.95$\pm$0.03	&	 0.67$\pm$0.07     &	 0.64$^{+0.09}_{-0.06}$       &        4.26$\pm$0.12	           &	$-0.58\pm$0.08  	    &	      4.8$\pm$0.4 		    &		  \ldots        	     &        \ldots		   &	    \ldots				\\
1.06$\pm$0.03	&	 $<$0.5            &	 0.4  {\it (fixed)}	      &        4.04$\pm$0.13	           &	$-0.33\pm$0.05  	    &	      6.9$\pm$0.6		    &		  \ldots        	     &        \ldots		   &	    \ldots				\\
1.15$\pm$0.03	&	 0.60$\pm$0.05     &	 0.58$\pm$0.04  	      &        3.52$\pm$0.07	           &	$-0.42^{+0.04}_{-0.07}$     &         6.8$\pm$0.4		    &		  \ldots        	     &        \ldots		   &	    \ldots				\\
1.25$\pm$0.03	&	 $<$0.5 	   &	 0.5  {\it (fixed)}	      &        3.55$\pm$0.12	           &    $-0.156^{+0.014}_{-0.021}$  &         9.6$^{+0.4}_{-0.8}$           &		  \ldots        	     &        \ldots		   &	    \ldots				\\
1.35$\pm$0.03	&	 0.43$\pm$0.09     &	 0.60$\pm$0.15  	      &        3.24$\pm$0.11	           &	$-0.29\pm$0.10 		    &         9.2$\pm$1.2		    &		  \ldots        	     &        \ldots		   &	    \ldots				\\
1.45$\pm$0.04	&	 0.63$\pm$0.04     &	 0.68$^{+0.10}_{-0.06}$       &        \ldots   	           &	    \ldots		    &	     \ldots		     	    &		  \ldots        	     &        \ldots		   &	    \ldots				\\
1.55$\pm$0.03	&	 0.71$\pm$0.04     &	 0.54$\pm$0.03  	      &        \ldots                      &	    \ldots		    &	     \ldots		     	    &		  \ldots        	     &        \ldots		   &	    \ldots				\\
1.65$\pm$0.04	&	 0.62$\pm$0.05     &	 1.29$\pm$0.19  	      &        \ldots                      &	    \ldots		    &	     \ldots		     	    &		  2.2$\pm$0.3 		     &        21$\pm$5  	   &	    10$\pm$2			        \\
1.76$\pm$0.04	&	 0.62$\pm$0.04     &	 1.6 $\pm$0.2		      &        \ldots                      &	    \ldots		    &	     \ldots		     	    &		  2.55$\pm$0.13		     &        15.4$\pm$1.8	   &	    9.2$\pm$0.7 			\\
1.86$\pm$0.04	&	 0.70$\pm$0.03     &	 1.48$\pm$0.14  	      &        \ldots                      &	    \ldots		    &	     \ldots		     	    &		  2.63$\pm$0.12		     &        13.5$\pm$1.3	   &	    9.9$\pm$0.6 			\\
1.95$\pm$0.04	&	 0.82$\pm$0.03     &	 0.92$\pm$0.10  	      &        \ldots                      &	    \ldots		    &	     \ldots		     	    &		  3.42$\pm$0.17		     &        9.4$\pm$0.3	   &	    9.6$\pm$0.3 			\\
2.04$\pm$0.04	&	 1.34$\pm$0.03     &	 0.77$\pm$0.03  	      &        \ldots                      &	    \ldots		    &	     \ldots		     	    &		  2.9$\pm$0.9 		     &        5.3$\pm$1.3	   &	    11.0$\pm$0.5			\\
2.13$\pm$0.05	&	 1.03$\pm$0.05     &	 1.01$\pm$0.11  	      &        \ldots                      &	    \ldots		    &	     \ldots		     	    &		  3.6$\pm$0.2 		     &        7.1$\pm$0.8	   &	    7.1$\pm$0.20			\\
2.23$\pm$0.06	&	 0.85$\pm$0.05     &	 0.86$^{+0.11}_{-0.07}$       &        \ldots                      &	    \ldots		    &	     \ldots		     	    &		  2.50$\pm$0.17		     &        6.0$\pm$0.9	   &	    5.8$\pm$0.2 			\\
2.34$\pm$0.05	&	 1.02$\pm$0.08     &	 1.12$^{+0.23}_{-0.16}$       &        \ldots         	           &	    \ldots		    &	     \ldots		     	    &		  2.4$^{+0.5}_{-0.3}$        &        9$^{+4}_{-2}$        &        5.6$\pm$0.7			        \\
2.45$\pm$0.05	&	 0.52$\pm$0.05     &	 3.93$^{+2.2}_{-1.1}$	      &        \ldots                      &	    \ldots		    &	     \ldots		     	    &		  2.6$^{+0.5}_{-0.3}$        &        12$^{+5}_{-3}$       &        3.9$^{+0.7}_{-1.0}$		        \\
2.54$\pm$0.04	&	 0.67$\pm$0.06     &	 2.5$\pm$0.4		      &        \ldots                      &	    \ldots		    &	     \ldots		     	    &		  2.7$\pm$0.2 		     &        9.0$\pm$1.6	   &	    4.8$\pm$0.4 			\\
2.65$\pm$0.04	&	 0.66$\pm$0.08     &	 1.7$\pm$0.4		      &        \ldots                      &	    \ldots		    &	     \ldots		     	    &		  3.5$\pm$0.3 		     &        16$\pm$3  	   &	    5.8$\pm$0.7 			\\
2.75$\pm$0.04	&	 0.59$\pm$0.07     &	 2.8$^{+0.9}_{-0.6}$	      &        \ldots                      &	    \ldots		    &	     \ldots		     	    &		  3.82$\pm$0.17		     &        19$\pm$3  	   &	    5.9$\pm$0.4 			\\
2.85$\pm$0.04	&	 0.46$\pm$0.06     &	 3.2$^{+1.5}_{-0.8}$	      &        \ldots                      &	    \ldots		    &	     \ldots		     	    &		  3.48$\pm$0.14		     &        14.7$\pm$1.6	   &	    5.3$\pm$0.4 			\\
2.95$\pm$0.04	&	 0.55$\pm$0.05     &	 2.7$\pm$0.5		      &        \ldots                      &	    \ldots		    &	     \ldots		     	    &		  3.66$\pm$0.11		     &        15.9$\pm$1.3	   &	    6.0$\pm$0.4 			\\
3.10$\pm$0.06	&	 0.64$\pm$0.03     &	 2.6$\pm$0.3		      &        \ldots                      &	    \ldots		    &	     \ldots		     	    &		  3.42$\pm$0.06		     &        14.1$\pm$0.7	   &	    6.1$\pm$0.2 			\\
3.29$\pm$0.07	&	 0.80$\pm$0.03     &	 2.12$\pm$0.18  	      &        \ldots                      &	    \ldots		    &	     \ldots		     	    &		  3.23$\pm$0.09		     &        16.3$\pm$1.1	   &	    6.3 $\pm$0.4			\\
3.50$\pm$0.07	&	 0.85$\pm$0.03     &	 2.14$\pm$0.16  	      &        \ldots                      &	    \ldots		    &	     \ldots		     	    &		  2.84$\pm$0.08		     &        17.4$\pm$1.3	   &	    7.0$\pm$0.5 			\\
3.70$\pm$0.07	&	 1.00$\pm$0.03     &	 1.83$\pm$0.10  	      &        \ldots                      &	    \ldots		    &	     \ldots		     	    &		  2.23$\pm$0.09		     &        15.7$\pm$1.7	   &	    9.5$\pm$0.5 			\\
3.89$\pm$0.06	&	 1.03$\pm$0.03     &	 1.90$\pm$0.10  	      &        \ldots                      &	    \ldots		    &	     \ldots		     	    &		  1.75$\pm$0.09		     &        12.6$\pm$1.7	   &	    10.7$\pm$0.7			\\
4.12$\pm$0.08	&	 1.03$\pm$0.03     &	 1.90$\pm$0.10  	      &        \ldots                      &	    \ldots		    &	     \ldots		     	    &		  1.49$^{+0.24}_{-0.15}$     &        23$^{+7}_{-5}$	    &	    7.0$^{+2.0}_{-3.5}$ 		\\
4.37$\pm$0.08	&	 1.22$\pm$0.03     &	 1.79$\pm$0.08  	      &        \ldots                      &	    \ldots		    &	     \ldots		     	    &		  1.40$^{+0.24}_{-0.16}$     &        23$^{+10}_{-7}$	   &	    9$^{+2}_{-3}$			\\
4.71$\pm$0.15	&	 1.65$\pm$0.03     &	 1.81$\pm$0.06  	      &        \ldots                      &	    \ldots		    &	     \ldots		     	    &		  0.94$\pm$ 0.12	     &        11 $\pm$3 	   &	   8 {\it (fixed)}		        \\
5.13$\pm$0.10	&	 1.61$\pm$0.04     &	 1.80$\pm$0.08  	      &        \ldots                      &	    \ldots		    &	     \ldots		     	    &		  $<0.6$    		     &        11 {\it (fixed)}     &	   8 {\it (fixed)}		        \\
\hline
\end{tabular}
\\
\noindent {\small $^a$ Integrated between 0.1 Hz and 1 Hz} \\   
\noindent {\small $^b$ Integrated between 1 Hz and 100 Hz} 
}
\caption{Fit results for the noise components at low frequencies.\label{tab:noise}}
\end{table}

\newpage
\clearpage

\pagestyle{plain}

\renewcommand{\arraystretch}{1.0}

\begin{table}[t]
\begin{center}
{\scriptsize
\begin{tabular}{cccccccc}
\hline
\hline
		& \multicolumn{3}{c}{Lower kHz QPO} & \multicolumn{3}{c}{Upper kHz QPO} &   \\
  $S_z$ 	& Rms (\%)		& FWHM (Hz)	      & Frequency (Hz)      & Rms (\%)  	    & FWHM (Hz) 	  & Frequency (Hz)     & Difference (Hz)      \\
\hline
$-0.32\pm$0.04  &  \ldots		&  \ldots	      &  \ldots 	    &  $<$7		    &  \raisebox{.0ex}[2.5ex][.0ex]{140$^{+134}_{-68}$} &  641$^{+35}_{-20}$ &   \ldots	      \\
$-0.23\pm$0.03  &  \ldots		&  \ldots	      &  \ldots 	    &  4.4$\pm$0.6	    &  100$^{+46}_{-28}$  &  637$\pm$9        &   \ldots	      \\
$-0.15\pm$0.03  &  \ldots		&  \ldots	      &  \ldots 	    &  5.4$^{+1.0}_{-0.6}$  &  192$^{+72}_{-45}$  &  618$\pm$12       &   \ldots	      \\
$-0.04\pm$0.03  &  \ldots		&  \ldots	      &  \ldots 	    &  5.7$\pm$0.4	    &  174$\pm$25	  &  650$\pm$7        &   \ldots	      \\
0.05$\pm$0.03   &  \ldots       	&  \ldots             &  \ldots  	    &  5.5$\pm$0.4 	    &  189$\pm$29 	  &  663$\pm$8        &   \ldots	      \\
0.16$\pm$0.03   &  \ldots       	&  \ldots             &  \ldots  	    &  5.4$\pm$0.4 	    &  178$\pm$26 	  &  678$\pm$7        &   \ldots	      \\
0.25$\pm$0.03   &  \ldots       	&  \ldots             &  \ldots  	    &  5.8$\pm$0.4 	    &  205$\pm$24 	  &  689$\pm$6        &   \ldots	      \\
0.35$\pm$0.04   &  \ldots       	&  \ldots             &  \ldots  	    &  4.8$\pm$0.3 	    &  175$\pm$21  	  &  722$\pm$6        &   \ldots	      \\
0.46$\pm$0.04   &  $<$2.0	       	&  130 {\it (fixed)}  &  475 {\it (fixed)}  &  4.6$\pm$0.2 	    &  164$\pm$16  	  &  759$\pm$5        &   \ldots	      \\
0.56$\pm$0.03   &  2.4$\pm$0.4  	&  145$\pm$55         &  517$\pm$18	    &  4.4$\pm$0.2 	    &  167$\pm$18 	  &  780$\pm$5        &   263$\pm$18          \\
0.65$\pm$0.03   &  2.7$^{+0.4}_{-0.3}$  &  137$^{+49}_{-30}$  &  519$^{+13}_{-9}$   &  4.46$\pm$0.20  	    &  160$\pm$15  	  &  804$\pm$4        &   284$\pm$11          \\
0.75$\pm$0.03   &  1.6$\pm$0.4  	&  85$\pm$59          &  530$\pm$20	    &  4.3$\pm$0.2 	    &  181$\pm$21   	  &  823$\pm$7        &   293$\pm$20          \\
0.85$\pm$0.03   &  2.4$\pm$0.3  	&  87$\pm$25  	      &  555$\pm$8	    &  4.0$\pm$0.3 	    &  154$\pm$20 	  &  849$\pm$6        &   294$\pm$10          \\
0.95$\pm$0.03   &  3.1$\pm$0.3  	&  104$\pm$25         &  607$\pm$8	    &  3.5$\pm$0.4 	    &  156$\pm$33 	  &  915$\pm$11       &   308$\pm$14          \\
1.06$\pm$0.03   &  3.6$\pm$0.3  	&  144$\pm$24         &  655$\pm$8	    &  3.5$\pm$0.3 	    &  149$\pm$30  	  &  950$\pm$8        &   294$\pm$11          \\
1.15$\pm$0.03   &  2.9$\pm$0.2  	&  87$\pm$15  	      &  682$\pm$5	    &  3.2$\pm$0.4 	    &  200$\pm$49  	  &  953$\pm$14       &   271$\pm$15          \\
1.25$\pm$0.03   &  3.0$^{+0.3}_{-0.2}$  &  112$^{+27}_{-8}$   &  714$\pm$7	    &  2.4$^{+0.4}_{-0.3}$  &  116$^{+49}_{- 33}$ &  973$\pm$11       &   258$\pm$13          \\
1.35$\pm$0.03   &  2.0$\pm$0.3  	&  67$\pm$24          &  752$\pm$8	    &  1.8$\pm$0.4 	    &  115$^{+106}_{-60}$ &  1011$\pm$24      &   259$\pm$25          \\
1.45$\pm$0.04   &  2.0$\pm$0.3  	&  98$\pm$41          &  794$\pm$14	    &  1.7$\pm$0.3 	    &  61$\pm$28  	  &  1033$\pm$9       &   239$\pm$17          \\
1.55$\pm$0.03   &  $<$2.3       	&  80$\pm$59 	      &  830$\pm$19 	    &  1.8$\pm$0.3 	    &  76$\pm$32 	  &  1063$\pm$13      &   \ldots	      \\
1.65$\pm$0.04   &  \ldots       	&  \ldots             &  \ldots 	    &  1.8$^{+0.5}_{-0.3}$  &  83$^{+66}_{-37}$   &  1087$\pm$15      &   \ldots	      \\
1.76$\pm$0.04   &  \ldots       	&  \ldots             &  \ldots 	    &  $<$1.5               &  90 {\it (fixed)}   &  1085 {\it (fixed)} &   \ldots	      \\
\hline
\end{tabular}
}
\end{center}
\caption{Fit results for the high frequency QPOs.\label{khz_tab}}
\end{table}

\end{document}